\newif \ifcomments \commentstrue
\newif \iffull \fulltrue
\newif \ifCCS \CCStrue
\newif \ifPETS \PETStrue

\def\cameraReady{} 

\fullfalse
\CCStrue
\PETSfalse

\iffull
\documentclass[11pt]{article}
\usepackage[utf8]{inputenc}
\usepackage[letterpaper,margin=1in,bmargin=1.5in]{geometry}
\usepackage[dvipsnames]{xcolor}
\fi

\ifCCS
  \PassOptionsToPackage{dvipsnames}{xcolor}
  \documentclass[sigconf]{acmart}
  \setcopyright{none}
  \acmConference[Anonymous Submission to ACM CCS 2022]{ACM Conference on Computer and Communications Security}{Due Jan 14 2022}{Los Angeles, U.S.A.}
  \acmYear{2022}

  \renewcommand\footnotetextcopyrightpermission[1]{}

  \makeatletter
    \renewcommand\paragraph{\@startsection{paragraph}{4}{\z@}%
      {-.5\baselineskip \@plus 0.2ex \@minus 0.1ex}
      {-3.5\p@}%
      {\ACM@NRadjust{\normalfont\normalsize\bfseries\boldmath\@adddotafter}}}
\makeatother
\fi

\ifPETS
    \PassOptionsToPackage{dvipsnames}{xcolor}
    \documentclass[USenglish,oneside,twocolumn]{article}

    \usepackage[utf8]{inputenc}
    \usepackage[big]{dgruyter_NEW}
    \usepackage{booktabs}

    \cclogo{\includegraphics{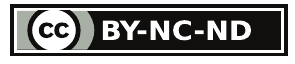}}

    \renewcommand{\paragraph}[1]{\vskip 1em \noindent\textbf{#1}}
\fi

\usepackage{setspace}
\usepackage{tikz}
\usetikzlibrary{calc,math, arrows,arrows.meta,shapes,positioning,matrix,decorations.pathreplacing}
\usepackage{amsfonts,amsmath,amsthm}
\usepackage{enumitem}
\usepackage{cryptocode} 

\ifPETS \else
\usepackage{multirow, ctable, tabu, makecell, graphics}
\usepackage{algorithm}
\usepackage[noend]{algpseudocode}
\fi

\usepackage[nameinlink,capitalize]{cleveref}
\usepackage{pifont}
\usepackage{xspace}
\usepackage{bm}

\iffull
\usepackage{amssymb}
\usepackage{authblk}
\usepackage[style=numeric-comp,backend=bibtex]{biblatex}
\usepackage[pdfauthor={},pdfpagelabels=true,linktocpage=true,hidelinks,linktoc=all,colorlinks]{hyperref}
\hypersetup{linkcolor = {black}, citecolor = {magenta}, urlcolor = {black}}
\fi

\ifPETS
\usepackage{afterpage}
\usepackage{multirow, ctable, tabu, makecell, graphics}
\usepackage{algorithm}
\usepackage[noend]{algpseudocode}
\makeatletter
\def\plist@algorithm{Alg.\space}
\makeatother
\fi

\ifCCS
\setlist{leftmargin=1.3em}
\fi

\newcommand{\sysname}{Zef\xspace}

\newcommand{\keyword}[1]{\normalfont \textsf{#1}}

\newcommand{\F}{\mathbb{F}}     

\newcommand{\len}[1]{\left|#1\right|}
\newcommand{\thtext}{{\rm th}}

\ifdefined\secparam\renewcommand{\secparam}{\lambda}\else\newcommand{\secparam}{\lambda}\fi

\newcommand{\sk}{{sk}}
\newcommand{\pubk}{\textnormal{\textsf{pk}}}
\newcommand{\pk}{{pk}}
\newcommand{\vk}{{vk}}

\newcommand{\rh}{{rh}}
\newcommand{\rs}{{rs}}
\newcommand{\rk}{{rk}}
\renewcommand{\rq}{{rq}}
\newcommand{\rv}{{rv}}

\newcommand{\ck}{{ck}}
\newcommand{\cq}{{cq}}
\newcommand{\cv}{{cv}}

\newcommand{\id}{{id}}

\newcommand{\com}{\textsf{com}}
\newcommand{\cm}{{cm}}

\newcommand{\cert}{\keyword{cert}}

\newcommand{\val}{V}
\newcommand{\amount}{\keyword{amount}}

\newcommand{\nextsequence}{\keyword{next\_sequence}}
\newcommand{\confirmed}{\keyword{confirmed}}
\newcommand{\recv}{\keyword{received}}  
\newcommand{\pending}{\keyword{pending}}

\newcommand{\hash}{\keyword{hash}}
\newcommand{\bal}{\keyword{balance}} 
\newcommand{\spent}{\keyword{spent}}
\newcommand{\parent}{\keyword{parent}}

\newcommand{\blind}{\keyword{blind}}

\newcommand{\unblind}{\keyword{unblind}}

\newcommand{\true}{\keyword{true}}

\ifcomments
    \newcommand{\mahimna}[1]{\textsf{\color{blue}{[Mahimna: {#1}]}}}
    \newcommand{\alberto}[1]{\textsf{\color{violet}{[Alberto: {#1}]}}}
    \newcommand{\mathieu}[1]{\textsf{\color{olive}{[Mathieu: {#1}]}}}
    \newcommand{\george}[1]{\textsf{\color{orange}{[George: {#1}]}}}
    \newcommand{\todo}[1]{\textsf{\color{red}{[Todo: {#1}]}}}

\else
    \newcommand{\mahimna}[1]{}
    \newcommand{\alberto}[1]{}
    \newcommand{\mathieu}[1]{}
    \newcommand{\george}[1]{}
    \newcommand{\todo}[1]{}
\fi

\theoremstyle{definition}

\theoremstyle{remark}

\newcommand{\mypara}{\paragraph}

\newcommand*\circled[1]{\tikz[baseline=(char.base)]{
            \node[scale=0.8,shape=circle,fill=black!75,inner sep=2pt] (char) {\textcolor{white}{\textbf{#1}}};}}

\newcommand{\resizeboxcol}[2]{\iffull {#2} \else \resizebox{#1}{!}{#2} \fi}

\newcommand{\twocolsloppy}{\iffull\else\sloppy\fi}

\iffull  \fi

\begin{document}
\title{\sysname: Low-latency, Scalable, Private Payments}

\ifdefined\cameraReady

\ifCCS
  
\author{Mathieu Baudet}
\email{mathieu.baudet@zefchain.com}
\affiliation{Zefchain Labs*\thanks{*The main part of this work was conducted while the author was at Facebook.}}

\author{Alberto Sonnino}
\email{alberto@mystenlabs.com}
\affiliation{Mysten Labs*}

\author{Mahimna Kelkar}
\email{mahimna@cs.cornell.edu}
\affiliation{Cornell University}

\author{George Danezis}
\email{george@mystenlabs.com}
\affiliation{Mysten Labs and University College London (UCL)}

\fi
\ifPETS
\author*[1]{Mathieu Baudet}
\author[2]{Alberto Sonnino}
\author[3]{Mahimna Kelkar}
\author[4]{George Danezis}

\affil[1]{Zefchain Labs, E-mail: mathieu.baudet@zefchain.com}
\affil[2]{Mysten Labs, E-mail: alberto@mystenlabs.com}
\affil[3]{Cornell University, E-mail: mahimna@cs.cornell.edu}
\affil[4]{Mysten Labs \& University College London (UCL), E-mail: george@mystenlabs.com}
\fi

\else
  \author{}
\fi

\date{}

\ifCCS
    \begin{abstract}
We introduce \sysname, the first Byzantine-Fault Tolerant (BFT) protocol to support payments in anonymous digital coins at arbitrary scale. \sysname follows the communication and security model of FastPay~\cite{fastpay}: both protocols are asynchronous, low-latency, linearly-scalable, and powered by partially-trusted sharded authorities. \sysname further introduces opaque coins represented as off-chain certificates that are bound to user accounts. In order to hide the face values of coins when a payment operation consumes or creates them, \sysname uses random commitments and NIZK proofs. Created coins are made unlinkable using the blind and randomizable threshold anonymous credentials of Coconut~\cite{coconut}. To control storage costs associated with coin replay prevention, \sysname accounts are designed so that data can be safely removed once an account is deactivated. Besides the specifications and a detailed analysis of the protocol, we are making available an open-source implementation of \sysname in Rust. Our extensive benchmarks on AWS confirm textbook linear scalability and demonstrate a confirmation time under one second at nominal capacity. Compared to existing anonymous payment systems based on a blockchain~\cite{zcash,monero}, this represents a latency speedup of three orders of magnitude, with no theoretical limit on throughput.
\end{abstract}
    \maketitle
    \pagestyle{plain}
\fi


\section{Introduction}

Anonymous payment systems have been an exciting research area in cryptography since Chaum's seminal work~\cite{chaum1982ecash} on e-cash. Early e-cash schemes~\cite{chaum1982ecash,sander1999ecash,camenisch2005ecash} however required a centralized issuer to operate, usually in the form of a trusted commercial bank, which hampered their adoption.  In recent years, the advent of networks like Bitcoin has sparked renewed interest in privacy-preserving decentralized payment systems. A number of protocols~\cite{zerocoin,Ben-SassonCG0MTV14,monero} focusing on anonymous payments are now deployed as permissionless blockchains.

Compared to traditional global payment infrastructures (aka. RTGS systems~\cite{rtgs}), however, decentralized anonymous payment systems have not yet reached performance levels able to sustain large-scale adoption. For instance, due to high computational costs, only 2\% of Zcash~\cite{zcash} transactions commonly take advantage of the privacy features offered by the platform~\cite{zchainexplorer}.

At the other end of the performance spectrum, the FastPay protocol~\cite{fastpay} does not support anonymous payments but offers low-latency transfers in the range of 100-200 ms and arbitrary (linear) scalability. FastPay operates in the Byzantine-Fault-Tolerant~(BFT) model with an asynchronous network.
This makes FastPay suitable for a deployment as a high-performance sidechain of an existing blockchain.
Remarkably, in order to scale linearly, FastPay is built solely on consistent broadcast between validators---as opposed to using a BFT consensus (see e.g.,~\cite{cachinBook}).

In this work, we revisit the FastPay design with privacy, storage costs, and extensibility in mind. In effect, we propose \sysname, the first linearly-scalable BFT protocol for anonymous payments with sub-second confirmation time.

\paragraph{The \sysname Protocol.} \sysname extends FastPay with digital coins that are both opaque and unlinkable (in short \emph{anonymous}). To this aim, \sysname combines several privacy-preserving techniques: (i)~randomized commitments and Non-Interactive Zero-Knowledge (NIZK) proofs (e.g.,~\cite{zerocoin}) provide \emph{opacity}, that is, hide payment values; (ii)~blind and randomizable signatures (e.g.,~\cite{coconut}) ensure \emph{unlinkability}, meaning that the relation between senders and receivers is hidden.

\paragraph{Technical Challenges.} As FastPay, \sysname achieves linear scalability by relying only on consistent broadcast~\cite{cachinBook}. Implementing anonymous coins in this setting poses three important challenges.

\begin{itemize}
\item \textbf{Double spending:} In the absence of a consensus protocol between validators, one cannot track the coins that have been spent in a single replicated data-structure. When coins are consumed to create new ones, we must also ensure that intermediate messages cannot be replayed to mint a different set of coins. We address this challenge by tracking input coins in one \emph{spent list} per account and by introducing hash commitments to bind input coins with their outputs.

\item \textbf{Storage costs:} Maintaining a spent list for each account while sustaining high throughput raises the question of storage costs. Spent lists must be readily accessible thus cannot be stored in cold storage. To make things worse, user accounts in FastPay can never be deleted due to the risk of replay attacks.
%
To address this challenge, we design \sysname accounts so that account data are safely removable once an account is deactivated by its owner. Concretely, this requires changing how user accounts are addressed in the system: instead of public keys chosen by users, \sysname must generate a unique (i.e. non-replayable) address when a new account is created. However, in the absence of consensus, address generation cannot rely on a replicated state.

\item \textbf{Implementation of privacy primitives:} While creating NIZK proofs on a predicate involving blind signatures, value conservation, and range constraints is theoretically possible, we wish to avoid the corresponding engineering and computational complexity in our implementation. To do so, we combine the Coconut scheme~\cite{coconut} and Bulletproofs~\cite{bulletproofs} to implement digital coins directly.
\end{itemize}

\paragraph{Contributions.} (1)~To support digital coins while controlling storage costs, we revisit the design of FastPay accounts: we propose a unified protocol for scalable accounts operations where accounts are addressed by unique, non-replayable identifiers (UIDs) and support a variety of operations such as account creation, deactivation, transparent payments, and ownership transfer. Importantly, all account operations in \sysname, including generation of system-wide unique identifiers, are linearly scalable, consensus-free, and only require elementary cryptography (hashing and signing).
(2)~Building on these new foundations, we describe and analyze the first asynchronous BFT protocol for opaque, unlinkable payments with linear (aka ``horizontal") scalability and sub-second latency.
(3)~Finally, we are making available an open-source prototype implementation of \sysname in Rust and provide extensive benchmarks to evaluate both the scalability and the latency of anonymous payments.


\section{Background and Related Work}
\label{sec:background}

\paragraph{FastPay.} FastPay~\cite{fastpay} was recently proposed as a sidechain protocol for low-latency, high-throughput payments in the Byzantine-Fault Tolerant model with asynchronous communication.
\begin{itemize}
    \item \textbf{Sidechain protocol:} FastPay is primarily meant as a scalability solution on top of an existing blockchain with smart contracts (e.g. Ethereum~\cite{wood2014ethereum}).
    \item \textbf{Byzantine-Fault Tolerance:} $N=3f+1$ replicas called \emph{authorities} are designated to operate the system and process the clients' requests. A fixed set of at most $f$ authorities may be \emph{malicious} (i.e. deviate from the protocol).
    \item \textbf{Low latency:} Authorities do not interact with each other (e.g. running a mempool or a consensus protocol). Client operations succeed predictably after a limited number of client/authorities round trips. Notably, in FastPay, a single round-trip with authorities suffices to both initiate a payment and obtain a certificate proving that the transfer is final.
    \item \textbf{Scalability:} 
    Each authority operates an arbitrary number of logical shards, across many physical hosts. By design, each client request is processed by a single shard within each authority. Within an authority, communication between shards is minimal and never blocks a client request.
    \item \textbf{Asynchronous communication:} Malicious nodes may collude with the network to prioritize or delay certain messages. Progress is guaranteed when messages eventually arrive.
\end{itemize}

In a nutshell, the state of the Fastpay accounts is replicated on a set of authorities. Each account contains a public key that can authorize payments out, a sequence number and a balance. Account owners authorize payments by signing them with their account key and including the recipient amount and payment value. An authorized payment is sent to all authorities, who countersign it if it contains the next sequence number; there are enough funds; and, it is the first for this account and sequence number. A large enough number (to achieve quorum intersection) of signatures constitutes a certificate for the payment. Obtaining a certificate ensures the payment can eventually be executed (finality). Anyone may submit the certificate to the authorities that check it and update the sender account and recipient balance.

FastPay does not rely on State-Machine Replication (SMR) in the sense that it does not require authorities to agree on a single global state---as one could expect from a traditional sidechain. Doing so, FastPay avoids the end-to-end latency cost of gathering, disseminating, and executing large blocks of transactions, a de-facto requirement for high throughput with SMR solutions~\cite{hotstuff, mir-bft, dumbo,  narwhal-and-tusk}.

Despite the benefits listed above, until now, the FastPay protocol has been limited to transparent payments, that is, without any privacy guarantees. In fact, to ensure fund availability in worst-case scenarios, FastPay requires all past money transfers to be publicly available in clear text. This contrasts negatively with traditional retail payments (e.g. credit cards) where individual transactions remain within a private banking network.
Another technical limitation of FastPay is that unused accounts cannot be deleted. In a privacy-sensitive setting where users would never re-use the same account twice, this means that storage cost of authorities would grow linearly with the number of past transactions.

\paragraph{Existing private payment schemes.} Compared to payment channels (e.g.~\cite{lightning}), safety in FastPay and Zef does not require any upper bound on network delays and clients to stay connected (aka. a \emph{synchrony} assumption~\cite{dwork1988consensus}). Furthermore, the reliability of the Lighting Network~\cite{lightning} depends on the existence of pairwise channels, with the success of a payment between two random nodes being at most 70\%\cite{diar-lightning}. In contrast, coins delegated to a FastPay instance are always immediately transferable to any recipient that possesses a public key (resp. an account identifier in \sysname).

Several privacy-preserving payment systems have been proposed in the past, each based on a blockchain consensus and therefore not linearly scalable: Zcash, based on Zerocash~\cite{Ben-SassonCG0MTV14}, uses a zero-knowledge proof of set inclusion which is expensive to compute instead of an efficient threshold issuance credential scheme. As a result most transactions are unshielded, leading to a degradation in privacy~\cite{KapposYMM18}. Monero~\cite{monero} uses ring signatures to ensure transactions benefit from a small anonymity set. However, intersections attacks and other transaction tracing heuristics are applicable. This results in an uneven degree of privacy~\cite{MoserSHLHSHHMNC18}.


\section{Overview}
\label{sec:overview}

We present \sysname, an evolution of FastPay~\cite{fastpay} designed to support high-volume, low-latency payments, both anonymous and transparent, on top of a \emph{primary blockchain}. To do so, \sysname introduces a new notion of accounts, indexed by a unique identifier (UID) so that deactivated accounts can have their data safely removed.

\mypara{Authorities and quorums.}
We assume a primary blockchain which supports smart contracts (e.g., Ethereum~\cite{wood2014ethereum}). In a typical deployment, we expect \sysname to be ``pegged'' to the primary chain through a smart contract, thereby allowing transfers of assets in either direction~\cite{pegged-sidechains}. The \sysname smart contract holds the reserve of assets (e.g., coins) and delegates their management to a set of external nodes called \emph{authorities}
. For brevity, in the rest of this paper, we focus on the \sysname system and omit the description of transfers between the primary blockchain and \sysname. The mechanics of such transfers is similar to "funding" and "redeeming" operations in FastPay~\cite{fastpay}.

\sysname is meant to be \emph{Byzantine-Fault Tolerant (BFT)}, that is, tolerate a subset of authorities that deviate arbitrarily from the protocol. We assume an \emph{asynchronous} network that may collude with malicious authorities to deliver messages in arbitrary order. The protocol makes progress when message are eventually delivered.

We assume that authorities have shared knowledge of each other's signing public keys. Each authority is also assigned a \emph{voting power}, which indicates how much control the authority has within the system. $N$ denotes the total voting power, while $f$ denotes the power held by adversarial authorities. In the simplest setting where each authority has a voting power of $1$ unit, $N$ denotes the total number of authorities and $f$ denotes the number of adversarial authorities tolerated by the system. In general, unequal voting powers may be used to reflect different \emph{stakes} locked by the authorities on the main blockchain. Similar to standard protocols, we require $0 \leq f < \frac{N}{3}$. The system parameters $N$ and $f$, as well as the public key and voting power of each authority are included in the \sysname smart contract during setup.

We use the word \emph{quorum} to refer to a set of signatures by authorities with a combined voting power of at least $N-f$. An important property of quorums, called \emph{quorum intersection}, is that for any two quorums, there exists an honest authority~$\alpha$ that is present in both.


\mypara{Cryptographic primitives.}
We assume a collision-resistant hash function, noted $\hash(\cdot)$, as well as a secure public-key signature scheme.
Informally, a \emph{random commitment} $\cm = \com_r(v)$ is an expression that provides a commitment over the value~$v$ (in particular, is collision-resistant) without revealing any information on~$v$, as long as the random seed~$r$ is kept secret.
A signing scheme supports \emph{blinding} and \emph{unblinding} operations iff~(i) a signature of a \emph{blinded message} $B = \blind(M, u)$ with \emph{blinding factor}~$u$ can be turned into a valid signature of $M$ by computing the expression $\unblind(B, u)$, and (ii)~provided that $u$ is a secret random value, an attacker observing~$B$ learns no information on~$M$.

Blind signatures will be used for anonymous coins in Section~\ref{sec:payments} together with an abstract notion of Non-Interactive Zero-Knowledge (NIZK) proof of knowledge.
We will also further assume that a public key $\pk_{\mathsf{all}}$ is set up between authorities in such a way that any quorum of signatures on $M$ may be aggregated into a single, secure \emph{threshold signature} of $M$, verifiable with $\pk_{\mathsf{all}}$. (See Appendix~\ref{sec:nizk_protocol} for a concrete instantiation.)

\mypara{Clients, requests, certificates, and coins.}
Clients to the \sysname protocol are assumed to know the public configuration of the system (see above) including networking addresses of authorities. Network interactions are always initiated by a client request. We distinguish \emph{account-based} requests, i.e., those targeting a specific account, noted~$R$, from \emph{free requests}~$R^*$. In what follows, all requests are account-based unless mentioned otherwise. Free requests will be used for coin creation in Section~\ref{sec:payments}.

\begin{figure}[t]
  \centering
  \resizeboxcol{\columnwidth}{
    \begin{tikzpicture}
      \node[draw,circle, minimum size=2.25cm] (0,0) {\textbf{\parbox{1.5cm}{\centering\small Account Owner}}};
      \node[draw,circle, minimum size=2.25cm] at (3.6,-3.5) {\textbf{\parbox{1.8cm}{\centering\small Recipients (if~any)}}};

      \tikzmath{\sidelen=3;}
      \tikzmath{\s=\sidelen/10;\inlen=\sidelen/3;}
      \begin{scope}[shift={(4.5,1.5)}]
        \draw[dashed] (0,0) rectangle (\sidelen,-1*\sidelen);
        \node[above] at (\sidelen/2,0) {\textbf{\sysname Committee}};

        \draw (\s,{-1*\s} ) rectangle ({\s+\inlen},{-1*(\s+\inlen)} );
        \draw ({\sidelen - \s - \inlen},{-1*\s} ) rectangle ({\sidelen-\s},{-1*(\s+\inlen)} );
        \draw (\s,{-1*(\sidelen - \s - \inlen)}) rectangle ({\s+\inlen},{-1*(\sidelen-\s)});
        \draw ({\sidelen - \s - \inlen},{-1*(\sidelen - \s - \inlen)}) rectangle ({\sidelen-\s},{-1*(\sidelen-\s)});
      \end{scope}

      \draw[solid,-latex, line width=0.25mm] (1,0.5) -- (4.5,0.5) node[midway,above]{\circled{1} Request $R$};
      \draw[solid,-latex, line width=0.25mm] (4.5,-0.5) -- (1,-0.5)  node[midway,above]{\circled{3} Vote on $R$};
      \draw[solid,-latex, line width=0.25mm] (0.2,-1.12) -- (0.2,-1.4) -- (4.5,-1.4) node[midway,above]{\circled{4} Confirm $\cert[R]$};
      \draw[solid,-latex, line width=0.25mm] (4.9,-3.5) -- (6,-3.5) -- (6,-1.5) node[midway,right]{\circled{4} Confirm $\cert[R]$};

      \begin{scope}[shift={(4.5,1.5)}]
        \draw[thick, -latex, shorten >=1pt] (\sidelen,{-1*(\s+\inlen/4)}) to [out=0,in=0,loop,looseness=6] (\sidelen,{-1*(\s+3*\inlen/4)}) node[right,shift={(1,\s)}] {\circled{2} Validate $R$};
        \draw[thick, -latex, shorten >=1pt] (\sidelen,{-1*(\sidelen - (\s+3*\inlen/4))}) to [out=0,in=0,loop,looseness=6] (\sidelen,{-1*(\sidelen - (\s+\inlen/4))}) node[right,shift={(1,\s)}] {\circled{5} Execute $R$};
      \end{scope}

      \draw[solid,-latex, line width=0.25mm] (0,-1.125) -- (0,-3.5) node[midway,right]{\circled{4} Show $\cert[R]$} -- (2.375,-3.5) ;

    \end{tikzpicture}
  }
  \caption{Request and execution of an account operation}
  \label{fig:account_requests}
\end{figure}
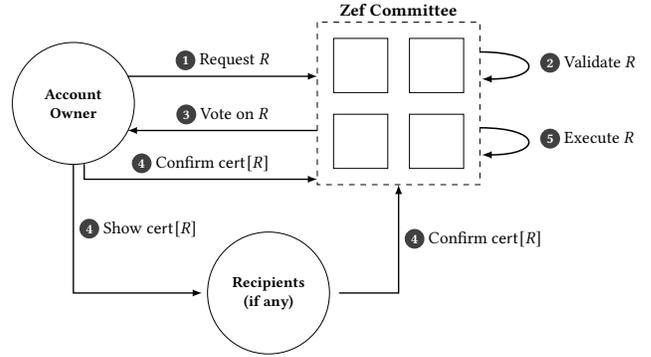

As illustrated in Figure~\ref{fig:account_requests}, clients may initiate a particular \emph{operation}~$O$ on an account that they own as follows: (i)~broadcast a request $R$ containing the operation $O$ and authenticated by the client's signature to the appropriate logical \emph{shard} of each authority~$\alpha$ ({\small\circled{1}}); and (ii)~wait for a quorum of responses, that is, sufficiently many answers so that the combined voting power of responding authorities reaches $N - f$.

An authority responds to a valid request $R$ by sending back a signature on~$R$, called a \emph{vote}, as acknowledgment~({\small\circled{3}}). After receiving votes from a quorum of authorities, a client forms a \emph{certificate}~$C$, that is, a request $R$ together with a quorum of signatures on $R$.
In the rest of this paper, we identify certificates on a same message~$M$ and simply write $C = \cert[M]$ when $C$ is a certificate on~$M$. Depending on the nature of~$M$ (e.g., anonymous coins in Section~\ref{sec:payments}) and implementation choices, the quorum of signatures in $C$ may be aggregated into a single threshold signature $\sigma$.

In addition to an operation~$O$, every request $R$ contains a \emph{sequence number} to distinguish successive requests on the same account. When a certificate $C = \cert[R]$ with the expected sequence number is received as a \emph{confirmation}~({\small\circled{4}}), this triggers the one-time execution of $O$~({\small\circled{5}}) and allows the user account behind $R$ to move on to the next sequence number. A confirmation certificate~$C$ also acts as a \emph{proof of finality}, that is, a verifiable document proving that the transaction (e.g., a payment) can be driven to success. In the case of payments, recipients should obtain and verify the certificate themselves before accepting the payment.

Finally, a third type of certificates associates a \emph{coin} to an account identifier~$\id$. Section~\ref{sec:payments} introduces \emph{anonymous coins} of the form $A = \cert[(\id, \cm)]$ for some appropriate commitment $\cm$ on the value $v$ of the coin.

\mypara{Accounts and unique identifiers.}
\sysname accounts are replicated across all authorities. For a given authority $\alpha$, we use the notation $X(\alpha)$ to denote the current view of $\alpha$ regarding some replicated data~$X$.
The features of \sysname accounts can be summarized as follows:
\begin{itemize}
    \item A \sysname account is addressed by an \emph{unique identifier} (UID or simply \emph{identifier} for short) designed to be non-replayable. We use $\id$, $\id_1$, \ldots to denote account identifiers. In practice, we expect users to publish the identifiers of some of their accounts, e.g. used for fund raising, and to keep other account identifiers secret to conceal their own payment activity---such as the timing and the number of opaque coins that they spend.
    \item Every operation executed on an account $\id$ follows from a certified request $C = \cert[R]$ that contains both $\id$ and a \emph{sequence number}~$n$. Validators must track the current sequence number of each account $\id$, so that operations on $\id$ are validated and executed in the natural order of sequence numbers $n = 0, 1, 2 \ldots$ Under BFT assumption, this ensures that all validators eventually execute the same sequence of operations on each account.
    \item To create a new account identified by $\id'$, the owner of an existing \emph{parent} account $\id$ must execute an account-creation operation. To ensure uniqueness, the new identifier is computed as the concatenation $\id' = \id :: n$ of the parent address $\id$ and the current sequence number $n$ of the parent account.
    \item Every account includes an optional public key $\pubk^\id(\alpha)$ to authenticate their owner, if any. When $\pubk^\id(\alpha) = \bot$, the account is said to be \emph{inactive}. \sysname makes it possible to safely and verifiably transfer the control of an account to another user by executing an operation to change the key $\pubk^\id(\alpha)$.
    \item In addition to the public balance, noted $\bal^\id(\alpha)$, the owner of an account $\id$ may possess a number of opaque coins $A = \cert[(\id, \cm)]$ cryptographically linked to the account. The face value of a coin is arbitrary and secretly encoded in $\cm$. Coins that are consumed (i.e. \emph{spent}) are tracked in a \emph{spent list} for each account---concretely, the protocol records $\cm$ in a set $\spent^\id(\alpha)$.
    \item An account can be \emph{deactivated} by setting $\pubk^\id(\alpha) = \bot$. This operation is final. Because identifiers $\id$ are never reused for new accounts, deactivated accounts may be safely deleted by authorities to reclaim storage (see discussion in Section~\ref{sec:accounts}).
\end{itemize}


\mypara{Sharding and cross-shard queries.}
In order to scale the processing of client requests, each \sysname authority~$\alpha$ may be physically divided in an arbitrary number of \emph{shards}. Every request $R$ sent to an account $\id$ in $\alpha$ is assigned a fixed shard as a public function of $\id$ and $\alpha$. If a request requires a modification of another \emph{target} account~$\id'$ (for instance, increasing its balance as part of a payment operation), the shard processing the confirmation of $R$ in $\alpha$ must issue an internal \emph{cross-shard} query to the shard of $\id'$. Cross-shard queries in \sysname are asynchronous messages within each authority. They are assumed to be perfectly reliable in the sense that are they are never dropped, duplicated, or tampered with.

\mypara{Transfer of anonymous coins.}
In \sysname, anonymous coins are both (i)~\emph{unlinkable} and (ii)~\emph{opaque} in the sense that during an anonymous payment: (i)~authorities cannot see or tracks users across coins being created; (ii)~authorities cannot see the values behind the commitments~$\cm$ of the coin being consumed or created.

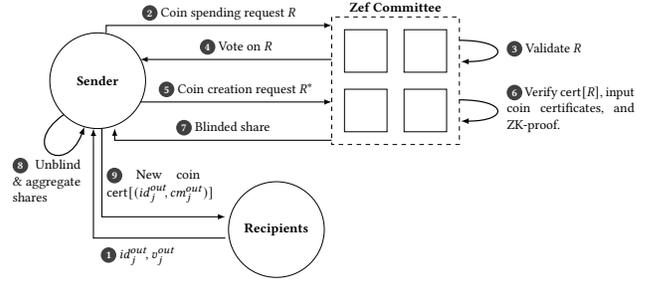
\begin{figure}
  \centering
  \resizeboxcol{\columnwidth}{
    \begin{tikzpicture}
      \node[draw,circle, minimum size=2.25cm] (0,0) {\textbf{Sender}};
      \node[draw,circle, minimum size=2.25cm] at (4.2,-3.5) {\textbf{Recipients}};

      \tikzmath{\sidelen=3;}
      \tikzmath{\s=\sidelen/10;\inlen=\sidelen/3;}
      \begin{scope}[shift={(5.5,1.5)}]
        \draw[dashed] (0,0) rectangle (\sidelen,-1*\sidelen);
        \node[above] at (\sidelen/2,0) {\textbf{\sysname Committee}};

        \draw (\s,{-1*\s} ) rectangle ({\s+\inlen},{-1*(\s+\inlen)} );
        \draw ({\sidelen - \s - \inlen},{-1*\s} ) rectangle ({\sidelen-\s},{-1*(\s+\inlen)} );
        \draw (\s,{-1*(\sidelen - \s - \inlen)}) rectangle ({\s+\inlen},{-1*(\sidelen-\s)});
        \draw ({\sidelen - \s - \inlen},{-1*(\sidelen - \s - \inlen)}) rectangle ({\sidelen-\s},{-1*(\sidelen-\s)});
      \end{scope}

      \draw[solid,-latex, line width=0.25mm] (0.2,1.12) -- (0.2,1.3) -- (5.5,1.3) node[midway,above]{\circled{2} Coin spending request $R$};
      \draw[solid,-latex, line width=0.25mm] (5.5,0.5) -- (1,0.5) node[midway,above]{\circled{4} Vote on $R$};
      \draw[solid,-latex, line width=0.25mm] (1,-0.5) -- (5.5,-0.5)  node[midway,above]{\circled{5} Coin creation request $R^*$};
      \draw[solid,-latex, line width=0.25mm] (5.5,-1.4) -- (0.4,-1.4) node[midway,above]{\circled{7} Blinded share} -- (0.4,-1.12);

      \begin{scope}[shift={(5.5,1.5)}]
        \draw[thick, -latex, shorten >=1pt] (\sidelen,{-1*(\s+\inlen/4)}) to [out=0,in=0,loop,looseness=6] (\sidelen,{-1*(\s+3*\inlen/4)}) node[right,shift={(1,\s)}] {\circled{3} Validate $R$};
        \draw[thick, -latex, shorten >=1pt] (\sidelen,{-1*(\sidelen - (\s+3*\inlen/4))}) to [out=0,in=0,loop,looseness=6] (\sidelen,{-1*(\sidelen - (\s+\inlen/4))}) node[right,shift={(1,\s)}] {\parbox{3cm}{\circled{6} Verify $\cert[R]$, input coin certificates, and ZK-proof.}};
      \end{scope}

      \draw[thick, -latex, shorten >=1pt] (-0.8,-0.8) to [out=-145,in=-110,loop,looseness=6] (-0.3,-1.05) node[below,shift={(-0.9,-0.65)}] {\parbox{1.6cm}{\circled{8} Unblind \& aggregate shares}};

      \draw[solid,-latex, line width=0.25mm] (0.1,-1.125) -- (0.1,-3.2) node[midway,right,shift={(0,-0.3)}]{\parbox{2.2cm}{\circled{9} New coin $\cert[(\id^{out}_j\!, \cm^{out}_j)]$}} -- (3,-3.2) ;

      \draw[solid,-latex, line width=0.25mm] (3,-3.7) -- (-0.1,-3.7) node[midway,below,shift={(-0.5,-0.1)}]{\circled{1} $\id^{out}_j$\!, $v^{out}_j$} -- (-0.1,-1.125) ;

    \end{tikzpicture}
  }
  \caption{An anonymous payment}
  \label{fig:anonymous_payments}
\end{figure}

Specifically, as illustrated in Figure~\ref{fig:anonymous_payments}, the owner of an account $\id$ may spend an anonymous coin $A^{in} = \cert[(\id, \cm^{in})]$ linked to $\id$ and create new anonymous coins $A^{out}_j$ as follows, using two communication round-trips with validators ({\small\circled{2}}--{\small\circled{7}}):
\begin{itemize}
    \item Obtain the receiving accounts $\id^{out}_j$ and desired coin values $v^{out}_j$ from recipients~({\small\circled{1}}).
    \item Compute fresh random commitments~$\cm^{out}_j$ for $v^{out}_j$ and fresh blinded messages $B_j = \blind((\id^{out}_j, \cm^{out}_j), u_j)$.
    \item Using the knowledge of the seed $r^{in}$ and coin value $v^{in}$ behind the random commitment $\cm^{in}$, construct an NIZK proof $\pi$ that the $B_j$ are well-formed---in particular, that the values $v^{out}_j$ are non-negative and that $\sum_j v^{out}_j = v^{in}$.
    \item Broadcast a request $R$ to spend the coin $A^{in}$ from the account $\id$, including the hash of the proof $\pi$ and the public values $\cm^{in}$ and $B_j$~({\small\circled{2}}).
    \item Aggregate the responses from a quorum of authorities into a certificate $C = \cert[R]$.
    \item Broadcast a suitable request $R^*$ containing the proof~$\pi$ together with~$C$, the coins~$A^{in}$, and the blinded messages $B_j$~({\small\circled{5}}).
    \item Obtain signature shares from a quorum of authorities for each $B_j$~({\small\circled{7}}), then unblind and aggregate the signatures shares to form new coins $A^{out}_j = \cert[(\id^{out}_j, \cm^{out}_j)]$~({\small\circled{8}}).
    \item Communicate each new coin $A^{out}_j$, as well as its commitment seed and value, privately to the owner of $\id^{out}_j$~({\small\circled{9}}).
\end{itemize}

Section~\ref{sec:payments} further elaborates on the creation of coins from public balances $\bal^\id(\alpha)$ and supporting multiple source accounts. The \sysname protocol also supports the converse operation consisting in transferring private coins into a public balance. Appendix~\ref{sec:nizk_protocol} provides more details on an efficient cryptographic instantiation of blind signatures and NIZK proofs using the Coconut scheme~\cite{coconut,rial2022security}. For comparison, we also describe a simplified protocol for transparent coins (i.e., without blinding and ZK-proofs) in Appendix~\ref{sec:transparent_coins}.

\paragraph{Bootstrapping account generation.}
%
In \sysname, creating a new account requires interacting with the owner of an existing \emph{parent} account. New identifiers are derived by concatenating the identifier of such a parent account with its current sequence number. This derivation ensures that identifiers are unique---and ultimately accounts are removable---while avoiding the overhead and the complexity of distributed random coin generation (see e.g.,~\cite{Cachin00randomoracles}).

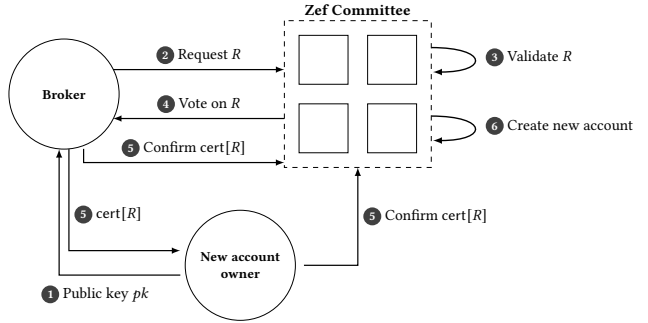
\begin{figure}[t]
  \centering
  \resizeboxcol{\columnwidth}{
    \begin{tikzpicture}
      \node[draw,circle, minimum size=2.25cm] (0,0) {\textbf{\parbox{1.5cm}{\centering\small Broker}}};
      \node[draw,circle, minimum size=2.25cm] at (3.6,-3.5) {\textbf{\parbox{1.8cm}{\centering\small New account owner}}};

      \tikzmath{\sidelen=3;}
      \tikzmath{\s=\sidelen/10;\inlen=\sidelen/3;}
      \begin{scope}[shift={(4.5,1.5)}]
        \draw[dashed] (0,0) rectangle (\sidelen,-1*\sidelen);
        \node[above] at (\sidelen/2,0) {\textbf{\sysname Committee}};

        \draw (\s,{-1*\s} ) rectangle ({\s+\inlen},{-1*(\s+\inlen)} );
        \draw ({\sidelen - \s - \inlen},{-1*\s} ) rectangle ({\sidelen-\s},{-1*(\s+\inlen)} );
        \draw (\s,{-1*(\sidelen - \s - \inlen)}) rectangle ({\s+\inlen},{-1*(\sidelen-\s)});
        \draw ({\sidelen - \s - \inlen},{-1*(\sidelen - \s - \inlen)}) rectangle ({\sidelen-\s},{-1*(\sidelen-\s)});
      \end{scope}

      \draw[solid,-latex, line width=0.25mm] (1,0.5) -- (4.5,0.5) node[midway,above]{\circled{2} Request $R$};
      \draw[solid,-latex, line width=0.25mm] (4.5,-0.5) -- (1,-0.5)  node[midway,above]{\circled{4} Vote on $R$};
      \draw[solid,-latex, line width=0.25mm] (0.4,-1.12) -- (0.4,-1.4) -- (4.5,-1.4) node[midway,above]{\circled{5} Confirm $\cert[R]$};
      \draw[solid,-latex, line width=0.25mm] (4.9,-3.5) -- (6,-3.5) -- (6,-1.5) node[midway,right]{\circled{5} Confirm $\cert[R]$};

      \begin{scope}[shift={(4.5,1.5)}]
        \draw[thick, -latex, shorten >=1pt] (\sidelen,{-1*(\s+\inlen/4)}) to [out=0,in=0,loop,looseness=6] (\sidelen,{-1*(\s+3*\inlen/4)}) node[right,shift={(1,\s)}] {\circled{3} Validate $R$};
        \draw[thick, -latex, shorten >=1pt] (\sidelen,{-1*(\sidelen - (\s+3*\inlen/4))}) to [out=0,in=0,loop,looseness=6] (\sidelen,{-1*(\sidelen - (\s+\inlen/4))}) node[right,shift={(1,\s)}] {\circled{6} Create new account};
      \end{scope}

      \draw[solid,-latex, line width=0.25mm] (0.1,-1.125) -- (0.1,-3.2) node[midway,right,shift={(0,-0.3)}]{\circled{5} $\cert[R]$} -- (2.375,-3.2) ;

      \draw[solid,-latex, line width=0.25mm] (2.375,-3.7) -- (-0.1,-3.7) node[midway,below,shift={(-0.5,-0.1)}]{\circled{1} Public key $\pk$} -- (-0.1,-1.125) ;

    \end{tikzpicture}
  }
  \caption{Request and creation of a new account}
  \label{fig:brokers}
\end{figure}

While \sysname lets any user derive new identifiers from an existing account that they possess, it is important that users can also obtain fresh identifiers in secret. Indeed, we expect users to regularly transfer the funds that they receive on public accounts into secret accounts in order to mitigate residual public information such as the timing and the number of coins spent from an account.

Therefore, for privacy reasons, we expect certain entities to specialize in creating fresh identifiers on behalf of other users. We call them \emph{brokers}. The role of brokers may also be assumed by authorities or delegated to third parties. In what follows, we assume that clients have a conventional way to pick an available broker and regularly create many identifiers for themselves ahead of time. The resulting interactions are summarized in Figure~\ref{fig:brokers}. To protect their identity, clients may also wish to interact with brokers and \sysname privately, say, using Tor~\cite{tor} or Nym\cite{nym}~({\small\circled{1}} and {\small\circled{5}}).

The fact that the role of broker can be delegated without risking account safety is an important property of the \sysname protocol discussed in Section~\ref{sec:accounts}. The solution relies on the notion of certificate for account operations---here used to prove finality of account creation, initialized with an authentication key chosen by the client.
In practical deployments, we expect authorities to charge a fee for account creation and brokers to forward this cost to their users plus a small margin. Discussing the appropriate pricing and means of payment is out of scope of this paper.

Finally, a \sysname system must be set up with a number of \emph{root} accounts (i.e., account without a parent). In the rest of the paper, we assume that the initial configuration of a \sysname system always includes one root account $\id_\alpha$ per authority~$\alpha$.

\mypara{Transfers of account ownership.} An interesting benefit of using unique identifiers as account addresses is that the authentication key $\pubk^\id(\alpha)$ can be changed. Importantly, the change of key can be certified to a new owner of the account. This unlocks a number of applications:
\begin{itemize}
    \item \textbf{Implicit transfer of coins}. Anonymous coins (see Section~\ref{sec:payments}) are defined as certificates of the form $A = \cert[(\id, \cm)]$ for some commitment value $\cm$. Spending $A$ to create new coins is an unlinkable operation but it reveals that a coin is spent from the account~$\id$. Account transfers provide an alternative way to transfer anonymous coins that is linkable but delays revealing the existence of coins.

    \item \textbf{Generalized authentication.} Account transfers opens the door to replacing $\pubk^\id(\alpha)$ with a variety of methods to authenticate a request made by the owner of an account. Common methods include multi-signatures, threshold signatures, and NIZK proofs of knowledge (see e.g.~\cite{bitcoin,zcash}).
    %

    \item \textbf{Lower account-generation latency.} While transferring ownership of an account~$\id$ requires the same number of messages as creating a new account, we will see in Section~\ref{sec:accounts} that it only involves executing an operation within the shard of $\id$ itself (i.e., no cross-shard requests are used). Hence, brokers who wish to provide new accounts with the lowest latency may create a pool of accounts in advance then re-assign UIDs to clients as needed.
\end{itemize}



\section{Account Management Protocol}
\label{sec:accounts}
We now describe the details of the \sysname protocol when it comes to account operations.
An upshot of our formalism is that it also naturally generalizes the FastPay transfer protocol~\cite{fastpay}. Notably, in \sysname, the one-time effect of a transaction consists in one of several possible operations, instead of transparent payments only. Additionally, in order to support deletion of accounts, \sysname must handle the fact that a recipient account might be deleted concurrently with a transfer.

\paragraph{Unique identifiers.}
A \emph{unique identifier} (UID or simply \emph{identifier}) is a non-empty sequence of numbers written as $\id = [n_1, \ldots, n_k]$ for some $1 \leq k \leq k_\mathsf{max}$. We use $::$ to denote the concatenation of one number at the end of a sequence: $[n_1, \ldots, n_{k+1}] = [n_1, \ldots, n_k]::n_{k+1}$ ($k < k_\mathsf{MAX}$). In this example, we say that $\id = [n_1, \ldots, n_k]$ is the \emph{parent} of $\id :: n_{k+1}$. We assume that every authority $\alpha$ possesses at least one \emph{root} identifier of length one: $\id_\alpha = [n_\alpha]$ such that the corresponding account is controlled by $\alpha$ at the initialization of the system (i.e., for every honest $\alpha'$, $\pubk^{\id_\alpha}(\alpha') = \alpha$).

\paragraph{Protocol messages.} A \emph{message} $\langle \mathsf{Tag}, \text{arg}_1, \ldots, \text{arg}_n \rangle$ is a sequence of values starting with a distinct marker $\mathsf{Tag}$ and meant to be sent over the network. In the remainder of the paper, we use capitalized names to distinguish message markers from mathematical functions (e.g. $\mathsf{hash}$) or data fields (e.g. $\pubk^\id(\alpha)$), and simply write $\mathsf{Tag}(\text{arg}_1, \ldots, \text{arg}_n)$ for a message.

\paragraph{Account operations.} An operation is a message $O$ meant to be executed once on a \emph{main} account $\id$, with possible effects on an optional \emph{recipient} account $\id'$. The operations supported by \sysname include the following messages:
\begin{itemize}
    \item $\mathsf{OpenAccount}(\id', \pk')$ to activate a new account with a fresh identifier $\id'$ and public key $\pk'$---possibly on behalf of another user who owns $\pk'$;
    \item $\mathsf{Transfer}(\id', \val)$ to transfer an amount of value $\val$ transparently to an account~$\id'$;
    \item $\mathsf{ChangeKey}(\pk')$ to transfer the ownership of an account;
    \item $\mathsf{CloseAccount}$ to deactivate the account~$\id$.
\end{itemize}
In Section~\ref{sec:payments}, we introduce two additional account operations $\mathsf{Spend}$ and $\mathsf{SpendAndTransfer}$.

\paragraph{Account states.}
Every authority $\alpha$ stores a map that contains the states of the accounts present in $\alpha$, indexed by their identifiers.
The state of the account~$\id$ includes the following data:
\begin{itemize}
    \item An optional public key $\pubk^\id(\alpha)$ registered to control $\id$, as seen before.
    \item A transparent (i.e., public) amount of value, noted $\bal^\id(\alpha)$ (initially equal to $\bal^\id(\mathsf{init})$, where $\bal^\id(\mathsf{init})$ is $0$ except for some special accounts created at the beginning).
    \item An integer value, written $\nextsequence^\id(\alpha)$, tracking the expected sequence number for the next operation on $\id$. (This value starts at $0$.)
    \item $\pending^\id(\alpha)$, an optional request indicating that an operation on $\id$ is pending confirmation (the initial value being $\bot$).
    \item A list of certificates, written $\confirmed^\id(\alpha)$, tracking all the certificates $C_n$ that have been confirmed by $\alpha$ for requests issued from the account $\id$. One such certificate is available for each sequence number~$n$ ($0 \leq n < \nextsequence^\id(\alpha)$).
    \item A second list of certificates, written $\recv^\id(\alpha)$, tracking all the certificates that have been confirmed by $\alpha$ and involving $\id$ as a recipient account.
\end{itemize}
In Section~\ref{sec:payments}, we will also assume a set of random commitments to track spent coins, noted $\spent^\id(\alpha)$.

\ifPETS \else
\begin{algorithm}[t]
\caption{Account operations (internal functions)}
\label{alg:account_operations}
\small
\begin{algorithmic}[1]
\algdef{SE}[ASYNC]{Async}{EndAsync}{\textbf{do asynchronously}}{\algorithmicend}%
\algtext*{EndAsync}%
\algnewcommand\algorithmicswitch{\textbf{switch}}
\algnewcommand\algorithmiccase{\textbf{case}}
\algnewcommand\algorithmicassert{\texttt{assert}}
\algnewcommand\Assert[1]{\State \algorithmicassert(#1)}%
\algdef{SE}[SWITCH]{Switch}{EndSwitch}[1]{\algorithmicswitch\ #1\ \algorithmicdo}{\algorithmicend\ \algorithmicswitch}%
\algdef{SE}[CASE]{Case}{EndCase}[1]{\algorithmiccase\ #1:}{\algorithmicend\ \algorithmiccase}%
\algtext*{EndSwitch}%
\algtext*{EndCase}%

\Function{Init}{$\id$} \Comment{Set up a new account if necessary}
\If {$\id \not\in \mathsf{accounts}$}
    \State $\pubk^\id \gets \bot$
    \State $\nextsequence^\id \gets 0$
    \State $\bal^\id \gets \bal^\id(\mathsf{init})$ \Comment{$0$ except for special accounts}
    \State $\confirmed^\id \gets [\,]$
    \State $\recv^\id \gets [\,]$
    \State $\spent^\id \gets \{\,\}$
\EndIf
\EndFunction
\Statex

\Function{ValidateOperation}{$\id$, $n$, $O$}
\Switch{$O$}
    \Case{$\mathsf{OpenAccount}(\id', \pk')$}
        \State ensure $\id' = \id :: \nextsequence^\id$
    \EndCase
    \Case {$\mathsf{Transfer}(\id', \val)$}
        \State ensure $0 < \val \leq \bal^\id$
    \EndCase
    \Case {$\mathsf{ChangeKey}(\pk') \;|\; \mathsf{CloseAccount}$}
        \State pass
    \EndCase
    \Case {$\mathsf{Spend}(\val, \cm, \sigma, h)$}
        \State ensure $0 \leq \val \leq \bal^\id$
        \State ensure $\cm \not\in \spent^\id$
        \State ensure $\sigma$ is a valid coin signature for $(\id, \cm)$
    \EndCase
    \Case {$\mathsf{SpendAndTransfer}(\id', \sigma, v, r)$}
        \State let $\cm = \com_{r}(v)$
        \State ensure $\cm \not\in \spent^\id$
        \State ensure $\sigma$ is a valid coin signature for $(\id, \cm)$
    \EndCase
\EndSwitch
\State \textbf{return} \true \Comment{$O$ is valid.}
\EndFunction

\Statex

\Function{ExecuteOperation}{$\id$, $O$, $C$}
\Switch{$O$}
    \Case{$\mathsf{OpenAccount}(\id', \pk')$}
        \color{BlueViolet}
        \Async \Comment{Cross-shard request to $\id'$}
            \State run \Call{Init}{$\id'$}
            \State $\pubk^{\id'} \gets \pk'$ \Comment{Activate authentication key}
            \State $\recv^{\id'} \gets \recv^{\id'} :: C$ \Comment{Update receiver's log}
        \EndAsync
        \color{black}
    \EndCase
    \Case {$\mathsf{Transfer}(\id', \val)$}
        \State $\bal^\id \gets \bal^\id - \val$ \Comment{Update sender's balance}
        \color{BlueViolet}
        \Async \Comment{Cross-shard request to $\id'$}
            \State run \Call{Init}{$\id'$}
            \State $\bal^{\id'} \gets \bal^{\id'} + \val$ \Comment{Receiver's balance}
            \State $\recv^{\id'} \gets \recv^{\id'} :: C$
        \EndAsync
        \color{black}
    \EndCase
    \Case{$\mathsf{ChangeKey}(\pk')$}
        \State $\pubk^\id \gets \pk'$ \Comment{Update authentication key}
    \EndCase
    \Case {$\mathsf{CloseAccount}$}
        \State $\pubk^\id \gets \bot$ \Comment{Make account inactive}
    \EndCase
    \Case {$\mathsf{Spend}(\val, \cm, \sigma, h)$}
        \State $\bal^\id \gets \bal^\id - \val$ \Comment{Update balance}
        \State $\spent^\id \gets \spent^\id \cup \{ cm \}$ \Comment{Mark coin as spent}
    \EndCase
    \Case {$\mathsf{SpendAndTransfer}(\id', \sigma, v, r)$}
        \State let $\cm = \com_{r}(v)$
        \State $\spent^\id \gets \spent^\id \cup \{ cm \}$ \Comment{Mark coin as spent}
        \color{BlueViolet}
        \Async \Comment{Cross-shard request to $\id'$}
            \State run \Call{Init}{$\id'$, $\bot$}
            \State $\bal^{\id'} \gets \bal^{\id'} + v$
            \State $\recv^{\id'} \gets \recv^{\id'} :: C$
        \EndAsync
        \color{black}
    \EndCase
\EndSwitch
\EndFunction
\end{algorithmic}

\end{algorithm}

\fi

\paragraph{Operation safety and execution.}
Importantly, account operations may require some validation before being accepted.
We say that an operation $O$ is \emph{safe} for the account $\id$ in~$\alpha$ if one of the following conditions holds:
\begin{itemize}
    \item $O = \mathsf{OpenAccount}(\id', \pk')$ and $\id' = \id :: \nextsequence^\id(\alpha)$;
    \item $O = \mathsf{Transfer}(\id', \val)$ and $0 \leq \val \leq \bal^\id(\alpha)$;
    \item $O = \mathsf{ChangeKey}(\pk')$ or $O = \mathsf{CloseAccount}$ (no additional verification).
\end{itemize}

When an operation $O$ for an account $\id$ is confirmed (i.e. a suitable certificate $C$ is received), we expect every authority~$\alpha$ to \emph{execute} the operation $O$ in following way:
\begin{itemize}
    \item if $O = \mathsf{OpenAccount}(\id', \pk')$, then the authority $\alpha$ uses a cross-shard request to set $\pubk^{\id'}(\alpha) = \pk'$; if necessary, a new account $\id'$ is created first;
    \item if $O = \mathsf{ChangeKey}(\pk')$, then the authority sets $\pubk^{\id}(\alpha) = \pk'$;
    \item if $O = \mathsf{Transfer}(\id', \val)$, the authority updates $\bal^{\id}(\alpha)$ by subtracting $\val$ and uses a cross-shard request to add $\val$ to $\bal^{\id'}(\alpha)$; if necessary, the account $\id'$ is created first using an empty public key $\pubk^{\id'}(\alpha) = \bot$;
    \item if $O = \mathsf{CloseAccount}$, then the authority deactivates the account by setting $\pubk^{\id}(\alpha) = \bot$.
\end{itemize}

These definitions translate to the pseudo-code in Algorithm~\ref{alg:account_operations}. The pseudo-code also includes the logging of certificates with $\confirmed^\id(\alpha)$ and $\recv^\id(\alpha)$ as well as additional operations $\mathsf{Spend}$ and $\mathsf{SpendAndTransfer}$ that will be described in Section~\ref{sec:payments}.

\paragraph{Account management protocol.} We can now describe the protocol steps for executing an operation~$O$ on an account~$\id$:
\begin{enumerate}
    \item A client knowing the signing key of $\id$ and the next sequence number $n$ signs a request $R = \mathsf{Execute}(\id, n, O)$ and broadcasts it to every authority in parallel, waiting for a quorum of responses.
    \item \label{step:vote} Upon receiving an authenticated request $R = \mathsf{Execute}(\id, n, O)$, an authority $\alpha$ must verify that $R$ is authenticated for the current account key $\pubk^\id(\alpha)$, that $\nextsequence^\id(\alpha) = n$, that the operation $O$ is safe (see above), and that $\pending^\id(\alpha) \in \{\bot, R\}$. Then, it sets $\pending^\id(\alpha) = R$ and returns a signature on $R$ to the client.
    \item \label{step:aggreg} The client aggregates signatures into a confirmation certificate $C = \cert[R]$.
    \item The client (or another stakeholder) broadcasts $\mathsf{Confirm}(C)$.
    \item \label{step:confirm} Upon receiving $\mathsf{Confirm}(C)$ for a valid certificate $C$ of value $R = \mathsf{Execute}(\id, n, O)$ when $O$ is an operation, each authority $\alpha$ verifies that
    $\pubk^\id(\alpha) \neq \bot$, $\nextsequence^\id(\alpha) = n$, then increments $\nextsequence^\id(\alpha)$, sets $\pending^\id(\alpha) = \bot$, adds $C$ to $\confirmed^\id(\alpha)$, and finally executes the operation $O$ once (see above).
\end{enumerate}

\ifPETS \else
\begin{algorithm}
\caption{Account service (message handlers)}
\label{alg:account_service}
\small
\begin{algorithmic}[1]
\Function{HandleRequest}{$\mathsf{auth}[R]$}
\State let $\mathsf{Execute}(\id, n, O) = R$
\State ensure $\pubk^\id \neq \bot$ \Comment{The account must be active}
\State verify that $\mathsf{auth}[R]$ is valid for $\pubk^\id$ \Comment{Check authentication}
\If {$\pending^\id \neq R$}
    \State ensure $\pending^\id = \bot$
    \State ensure $\nextsequence^\id = n$
    \State ensure \Call{ValidateOperation}{$\id$, $n$, $O$}
    \State $\pending^\id \gets R$ \Comment{Lock the account on $R$}
\EndIf
\State \textbf{return} $\Call{Vote}{R}$ \Comment{Success: return a signature of the request}
\EndFunction
\Statex
\Function{HandleConfirmation}{$C$}
\State verify that $C = \mathsf{cert}[R]$ is valid
\State let $\mathsf{Execute}(\id, n, O) = R$
\State ensure $\pubk^\id \neq \bot$ \Comment{Make sure the account is active}
\If {$\nextsequence^\id = n$}
    \State run \Call{ExecuteOperation}{$\id$, $O$, $C$}
    \State $\nextsequence^\id \gets n + 1$ \Comment{Update sequence number}
    \State $\pending^\id \gets \bot$ \Comment{Make the account available again}
    \State $\confirmed^\id \gets \confirmed^\id :: C$ \Comment{Log certificate}
\EndIf
\EndFunction
\end{algorithmic}
\end{algorithm}

\fi

The corresponding pseudo-code for the service provided by each authority $\alpha$ is summarized in Algorithm~\ref{alg:account_service}.
Importantly, \emph{inactive} accounts, i.e., those accounts~$\id$ satisfying $\pubk^\id(\alpha) = \bot$, cannot accept any request (step~(\ref{step:vote})) or execute any confirmed operation (step~(\ref{step:confirm})).
Note that step~(1) above implicitly assumes that all authorities are up-to-date with all past certificates. In practice, a client may need to provide each authority with missing confirmation certificates for past sequence numbers. (See also ``Liveness considerations" below.)

\paragraph{Agreement on account operations.} When it comes to the operations executed from one account~$\id$, the \sysname protocol guarantees that authorities execute the same sequence of operations in the same order.
%
Indeed, the quorum intersection property entails that two certificates $C$ and $C'$ must contain a vote by a same honest authority~$\alpha$. If they concern the same account $\id$ and sequence number~$n$, the verification by $\alpha$ in step~(\ref{step:vote}) above and the increment of $\nextsequence^\id(\alpha)$ in step~(\ref{step:confirm}) implies that $C$ and $C'$ certifies the same (safe) request $R$.

It is easy to see by induction on the length of $\id = [ n_1, \ldots, n_k]$ that each authority can only execute certified operations for a given $\id$ by following the natural sequence of sequence numbers (i.e., $\nextsequence^\id(\alpha) = 0, 1, \ldots$). Indeed, by the induction hypothesis (resp. by construction for the base case), at most one operation of the form $O = \mathsf{OpenAccount}(\id, ..)$ can ever be executed by $\alpha$ on the parent account of $\id$ (resp. as part of the initial setup if $\id$ has no parent). We also note that due to the checks in step~(\ref{step:confirm}), no operation can be executed from $\id$ while the account $\id$ is locally absent or if $\pubk^\id(\alpha) = \bot$. Account creation executed by the parent account of $\id$ is the only way for $\pubk^\id(\alpha)$ to be updated from an empty value $\bot$. Therefore, if an account $\id$ is deleted by $\alpha$ due to an operation $\mathsf{CloseAccount}$, it is necessarily so after $\mathsf{OpenAccount}(\id, ..)$ was already executed once. The account $\id$ may be created again by some operation $\mathsf{Transfer}(\id, \val)$ after deletion, but since $\mathsf{OpenAccount}(\id, ..)$ is no longer possible, $\pubk^\id(\alpha)$ will remain empty, thus no more operations will be executed from $\id$ at this point. Therefore, due to the checks in step~(\ref{step:confirm}), the operations executed on $\id$, necessarily while $\pubk^\id(\alpha) \neq \bot$, follows the natural sequence of sequence numbers.

\paragraph{Agreement on account states.} Let $\alpha$ be authority and $\id$ be an account such that $\pending(\alpha) = \bot$ and $\alpha$ has not executed an operation $\mathsf{CloseAccount}$ on $\id$ yet. We observe that the state of $\id$ seen by $\alpha$ is a deterministic function of the following elements:
\begin{itemize}
    \item the sequence of operations previously executed by $\alpha$ on $\id$, that is, the content of $\confirmed^\id(\alpha)$, and
    \item the (unordered) set of operations previously executed by $\alpha$ that caused a cross-shard request to $\id$ as recipient, that is, the content of $\recv^\id(\alpha)$.
\end{itemize}
Indeed, operations issued by $\id$ are of the form $\mathsf{ChangeKey}(\pk)$, $\mathsf{Transfer}(.., \val^{out}_j)$, and $\mathsf{OpenAccount}(..)$.
Similarly, possible operations received by $\id$ are of the form $\mathsf{OpenAccount}(\id, \pk)$ and $\mathsf{Transfer}(\id, \val^{in}_i)$.
We can determine the different components of the account $\id$ as seen by $\alpha$ as follows:
\begin{itemize}
    \item $\nextsequence^\id(\alpha)$ will be the size of $\confirmed^\id(\alpha)$;
    \item $\pubk^\id(\alpha)$ will be the last key set by $\mathsf{OpenAccount}(\id, \pk)$ (or an equivalent initial setup for special accounts) then subsequent $\mathsf{ChangeKey}(\pk)$ operations, and otherwise $\pubk^\id(\alpha) = \bot$;
    \item $\bal^\id(\alpha) =  \sum_i \val^{in}_i \,-\, \sum_j \val^{out}_j \,+\, \bal^\id(\mathsf{init})$, where $\bal^\id(\mathsf{init})$ denotes a possibly non-zero initial balance for some special accounts. (In the presentation of FastPay~\cite{fastpay}, additionally terms account for external transfers with the primary blockchain in replacement of $\bal^\id(\mathsf{init})$.)
\end{itemize}

\twocolsloppy The agreement property on account operations (see above) entails that whenever two honest authorities have executed the same operations, they must also agree on the current set of active accounts and their corresponding states. In other words, if for all $\id$,
$\confirmed^\id(\alpha) = \confirmed^\id(\alpha')$, then for all $\id$ such that $\pubk^\id(\alpha) \neq \bot$ or $\pubk^\id(\alpha') \neq \bot$, we have $\nextsequence^\id(\alpha) = \nextsequence^\id(\alpha')$, $\pubk^\id(\alpha) = \pubk^\id(\alpha')$, and $\bal^\id(\alpha) = \bal^\id(\alpha')$.

In particular, similar to the proof of FastPay~\cite{fastpay}, $\bal^\id(\alpha) \geq 0$ holds for every~$\id$ once every certified operations has been executed. Indeed, consider an honest authority which accepted to vote at step~(\ref{step:vote}) for the last transfer $\mathsf{Transfer}(.., \val^{out}_j)$ from $\id$.

\paragraph{Liveness considerations.}
\sysname guarantees that conforming clients may always (i)~initiate new valid operations on their active accounts and (ii)~confirm a valid certificate of interest as a sender or as a recipient.
We note that question~(i) is merely about ensuring that the sequence number of an active sender account can advance after a certificate is formed at step~(\ref{step:aggreg}). This reduces to the question~(ii) of successfully executing step~(\ref{step:confirm}) for any honest authority, given a valid certificate $C$.

If the client, an honest authority $\alpha$, or the network was recently faulty, it is possible that (a)~the sender account $\id$ may not be active yet at $\alpha$, or (b)~the sequence number of $\id$ may be lagging behind compared to the expected sequence number in $C$. In the latter case~(b), similarly to Fastpay, the client should replay the \emph{previously confirmed certificates} $C_i$ of the same account---defined as $C_i \in \confirmed^\id(\alpha')$ for some honest $\alpha'$---in order to bring an authority $\alpha$ to the latest sequence number and confirm $C$. In the case (a) where $\id$ is not active yet at $\alpha$, the client must confirm the creation certificate $C'$ of $\id$ issued by the parent account $\id' = \parent(id)$. This may recursively require confirming the history of $C'$. Note however that this history is still sequential (i.e. there is at most one parent per account) and the number of parent creation certificates is limited by $k_\mathsf{MAX}$.

Importantly, a certificate needs only be confirmed once per honest authority on behalf of all clients. Conforming clients who initiate transactions are expected to persist past certificates locally and pro-actively share them with all responsive authorities.

In practice, the procedure to bring authorities up-to-date can be implemented in a way that malicious authorities that would always request the entire history do not slow down the protocol. (See the discussion in FastPay~\cite{fastpay}, Section~5.)

\paragraph{Deactivation and deletion of accounts.} We have seen that once deactivated, an account~$\id$ plays no role in the protocol and that $\id$ will never be active again. Therefore, it is always safe for an authority to remove a deactivated account from its local storage.

This important result paves the way for \sysname deployments to control their storage cost by incentivizing users to regularly create new accounts and deactivate old ones. For instance, a deployment may limit the maximum sequence number for account operations and limit the number of opaque coins spent in each account.

Assuming that deactivated accounts are regularly produced, a simple strategy for an authority $\alpha$ to reclaim some local storage consists in deleting an account $\id$ whenever $\pubk^\id(\alpha)$ changes its value to $\bot$.
We note however that this strategy is only a best effort. Effectively reclaiming the maximum amount of storage available in the system requires addressing two questions:
\begin{enumerate}
    \item If an honest authority $\alpha$ deletes $\id$, how to guarantee that the account is not recreated later by $\alpha$;
    \item If an honest authority $\alpha$ deletes $\id$, how to guarantee that every other honest authority $\alpha' \neq \alpha$ eventually deletes $\id$.
\end{enumerate}

Regarding (1), when a cross-shard request is received for an operation $\mathsf{Transfer}(\id, \val)$, the current version of the protocol may indeed  re-create an empty account $\id$. This storage cost can be addressed by modifying \sysname so that an authority~$\alpha$ does not re-create~$\id$ (or quickly deletes it again) if it determines that no operation $O = \mathsf{OpenAccount}(\id, ..)$ can occur any more. This fact can be tested in background using $|\id| \leq k_\mathsf{max}$ cross-shard queries. Indeed, consider the opposite fact: an inactive account $\id = \id_0 :: n$ \emph{can become active} in $\alpha$ iff it holds that (i)~$\nextsequence^{\id_0}(\alpha) \leq n$ and (ii)~the parent account $\id_0$ is either active or can become active.

Regarding (2), we note that sending and receiving clients in payment operations have an incentive to fully disseminate the confirmation certificates to all authorities---rather than just a quorum of them---whenever possible. (The incentives are respectively to fully unlock the sender's account and to fully increase the receiver's balance in the eventuality of future unresponsive authorities.) However, such an incentive does not exist in the case of the $\mathsf{CloseAccount}$ operation. Therefore, in practical deployments of \sysname, we expect authorities to either communicate with each other a minima in background, or to incentivize clients to continuously disseminate missing certificates between authorities.

\paragraph{Security of account generation.} In the eventuality of malicious brokers, a client must always verify the following properties before using a new account~$\id'$:
\begin{itemize}
    \item The certificate $C$ returned by the broker is a valid certificate $C = \cert[R]$ such that $R = \mathsf{Execute}(\id, n, O)$ and $O = \mathsf{OpenAccount}(\id', \pk)$ for the expected public key $\pk$. (Under BFT assumption, this implies $\id' = \id :: n$.)
    \item If the client did not pick a fresh key $\pk$, it is important to also verify that $C$ is not being replayed.
\end{itemize}

For new accounts meant to be secret, clients should use a fresh public key $\pk$ and consider communicating with a broker privately (e.g. over Tor). How a client may anonymously purchase their first identifiers from a broker raises the interesting question of how to effectively bootstrap a fully anonymous payment system. (For instance, a certain number of fresh key-less accounts could be given away regularly for anyone to acquire and reconfigure them over Tor before receiving their very first anonymous payment.)
%

\mypara{Further comparison with FastPay.}
In FastPay, accounts are indexed by the public key~$\pk$ that controls payment transfers from the account. Such a key is also called a \emph{FastPay address}. The state of an account $\pk$ is replicated by every authority $\alpha$ and includes notably a balance $\mathsf{balance}^\pk(\alpha)$ and a sequence number $\mathsf{next\_sequence}^\pk(\alpha)$ used to prevent replay of payment certificates.

The definition of FastPay addresses entails that an account $\pk$ (even with balance $0$) can never be removed from the system. Indeed, after the information on the sequence number $\mathsf{next\_sequence}^\pk(\alpha)$ is lost, the account owner may re-create an account for the same public key $\pk$ and exploit $\mathsf{next\_sequence}^\pk(\alpha) = 0$ to replay all past transfers originating from $\pk$.
In a context of privacy-aware applications, users are less likely to re-use a same account $\pk$ many times, thus amplifying the storage impact of unused accounts.
While anonymous coins introduced next in Section~\ref{sec:overview} and~\ref{sec:payments} can easily be adapted to FastPay-like accounts indexed by $\pk$, this would cause requirements in local storage to never decrease even if some accounts were explicitly deactivated.

In \sysname, accounts are indexed by a unique identifier and deactivated accounts can be safely deleted. On the downside, new users must interact with a broker or an authority ahead of time to obtain fresh identifiers. Existing users may also choose to trade some privacy and derive identifiers from their existing account(s).

Cross-shard queries in both FastPay and \sysname are asynchronous in the sense that they do not block a client request to confirm a certificate (see Algorithm~\ref{alg:account_operations}). This is crucial to guarantee that an authority with a lagging view on a particular account can be brought up to date by providing missing certificate history for this account and its parents only---as opposed to exponentially many accounts. In \sysname, this property results from a careful design of the protocol allowing missing recipient accounts to be (re)created with an empty public key $\pubk^\id(\alpha) = \bot$ whenever needed. The uniqueness property of identifiers guarantees that a deleted account can never be reactivated later on.

\section{Anonymous Payments}
\label{sec:payments}
We now describe the \sysname protocol for anonymous payments using generic building blocks. In particular, we use a blind signature scheme, random commitments, and Zero-Knowledge (ZK) proofs in a black-box way. A more integrated realization of the protocol suitable for an efficient implementation is proposed in Appendix~\ref{sec:nizk_protocol}.

\mypara{Anonymous coins.}
An anonymous coin is a triplet $A = (\id, \cm, \sigma)$ where $\id$ is the unique identifier (UID) of an active account, $\cm$ is a random commitment on a value $v \in [0, v_\textsf{max}]$ using some randomness $r$, denoted $\cm = \com_r(v)$, and $\sigma$ is a threshold signature from a quorum of authorities on the pair $(\id, \cm)$.
Following the notations of Section~\ref{sec:overview}, an anonymous coin $A$ can also be seen as a certificate $A = \cert[(\id, \cm)]$.
To effectively \emph{own} a coin, a client must know the value $v$, the randomness $r$, and the secret key controlling $\id$.
To prevent double-spending, for every account $\id$, every authority keeps tracks of the coins that have already been spent by storing commitments $\cm$ in a set $\spent^\id(\alpha)$.

\mypara{Spending anonymous coins.}
We extend the account operations of Section~\ref{sec:accounts} with an operation $O = \mathsf{Spend}(\val, \cm, \sigma, h)$ meant to be included in a request $\mathsf{Execute}(\id, n, O)$. This operation prepares the creation of new coins by consuming one opaque coin $(\id, \cm, \sigma)$ and by transparently withdrawing some amount $\val$ from the account. The hash value $h$ forces the sender to commit to specific output coins (see next paragraph).
Following the framework of Section~\ref{sec:accounts}:
\begin{itemize}
    \item $O$ is \emph{safe} iff $\sigma$ is a valid signature for $(\id, \cm)$,  $0 \leq \val \leq \bal^\id(\alpha)$, and $\cm \not\in \spent^\id(\alpha)$.
    \item Upon receiving a valid certificate $C = \cert[R]$, the execution of $O$ consists in substracting $\val$ from $\bal^\id(\alpha)$ and adding $\cm$ to $\spent^\id(\alpha)$.
\end{itemize}
See algorithm~\ref{alg:account_operations} for the corresponding pseudo-code.

\mypara{Creating anonymous coins.}
Suppose that a user owns $\ell$ coins $A^{in}_i = (\id^{in}_i, \cm^{in}_i, \sigma^{in}_i)$ ($1 \leq i \leq \ell$) such that the $\cm^{in}_i$ are $\ell$ mutually distinct random commitments, and $\sigma^{in}_i$ is a coin signature on $(\id^{in}_i, \cm^{in}_i)$. Let $v^{in}_i$ be the value of the coin $A^{in}_i$. Let $\val^{in}_i \geq 0$ be a value that the user wishes to withdraw transparently from the account $\id^{in}_i$.
Importantly, we require commitments $\cm^{in}_i$ to be distinct but not the identifiers $\id^{in}_i$. This allows several coins to be spent from the same account.

We define the total input value of the transfer as $v = \sum_i v^{in}_i + \sum_i \val^{in}_i$.
To spend the coins into $d$ new coins with values $v^{out}_j$ ($1 \leq j \leq d$) such that $\sum_j v^{out}_j = v$, the sender requests a unique identifier $\id^{out}_j$ from each recipient, then proceeds as follows:
\begin{enumerate}
\item First, the sender constructs blinded messages $B_j$ and a zero-knowledge proof $\pi$ as follows:

\begin{enumerate}
\item For $1 \leq j \leq d$, sample randomness $r^{out}_j$ and set $\cm^{out}_j = \com_{r^{out}_j}(v^{out}_j)$.
\item For $1 \leq j \leq d$, sample random blinding factor $u_j$ and let $B_j = \blind((\id^{out}_j, \cm^{out}_j), u_j)$.

\item Construct a zero-knowledge proof $\pi$ for the following statement regarding $(\cm^{in}_1, \ldots, cm^{in}_\ell, \sum_i \val^{in}_i, B_1, \ldots, B_d)$: I know $v^{in}_i, r^{in}_i$ for each $1 \leq i \leq \ell$ and $v^{out}_j, r^{out}_j, u_j, \id^{out}_j$ for each $1 \leq j \leq d$ such that
\begin{itemize}
    \item $\cm^{in}_i = \com_{r^{in}_i}(v^{in}_i)$ and $\cm^{out}_j = \com_{r^{out}_j}(v^{out}_j)$
    \item $B_j = \blind((\id^{out}_j, \cm^{out}_j), u_j)$
    \item $\sum_i v^{in}_i + \sum_i \val^{in}_i = \sum_j v^{out}_j$
    \item Each value $v^{in}_i$ and $v^{out}_j$ is in $[0, v_\mathsf{max}]$
\end{itemize}
\end{enumerate}

\item For every input $i$, the sender obtains a certificate $C_i$ for the operation $O_i = \mathsf{Spend}(\val^{in}_i, \cm^{in}_i, \sigma^{in}_i, \hash(B_1, \ldots, B_d))$ then confirms $C_i$. Concretely, as detailed in Section~\ref{sec:accounts}, this means broadcasting an authenticated request $R_i = \mathsf{Execute}(\id^{in}_i, n_i, O_i)$ for a suitable sequence number $n_i$, obtaining a quorum of votes on $R_i$, then broadcasting $C_i = \cert[R_i]$.

\item Next, the sender broadcasts a free request $R^* = \mathsf{CreateAnonymousCoins}(\pi, C_1, \ldots, C_\ell, B_1, \ldots, B_d)$ and waits for a quorum of responses.

\item Upon receiving a free request of the form $R^* = \mathsf{CreateAnonymousCoins}(\pi, C_1, \ldots, C_\ell, B_1, \ldots, B_d)$ where
$C_i = \cert[R_i]$, $R_i = \mathsf{Execute}(\id_i, n_i, O_i)$, $O_i = \mathsf{Spend}(\val_i, \cm_i, \sigma_i, h_i$, each authority $\alpha$ verifies the following:
\begin{itemize}
    \item Every $C_i$ is a valid certificate for $R_i$. (Under BFT assumption, this implies that $\sigma_i$ is a valid signature on $(\id_i, \cm_i)$)
    \item The values $\cm_i$ are mutually distinct.
    \item $h_i = \hash(B_1, \ldots, B_d))$.
    \item The proof $\pi$ is valid for the public inputs $(\cm_1, \ldots, cm_\ell, \sum_i \val_i, B_1, \ldots, B_d)$.
\end{itemize}
The authority then responds with $d$ signature shares, one for each $B_j = \blind((\id^{out}_j, \cm^{out}_j); u_j)$.

\item For every $j$, the sender finally combines the signature shares received by a quorum of authorities, then uses $\unblind$ to obtain a signature $\sigma^{out}_j$ on $(\id^{out}_j, \cm^{out}_j$).

\item The $j^{th}$ recipient receives $(\id^{out}_j, \cm^{out}_j, v^{out}_j, r^{out}_j, \sigma^{out}_j)$. She verifies that the values and identifiers are as expected, that the commitments $\cm^{out}_j$ are mutually distinct and each
$\sigma^{out}_j$ is valid.

\end{enumerate}

We note that finality is achieved as soon as the request $R^*$ is formed by the sender.
The pseudo-code for coin creation is presented in Algorithm~\ref{alg:coin_creation}.

\ifPETS \else
\begin{algorithm}
\caption{Coin creation service}
\label{alg:coin_creation}
\begin{algorithmic}[1]
\algdef{SE}[ASYNC]{Async}{EndAsync}{\textbf{do asynchronously}}{\algorithmicend}%
\algtext*{EndAsync}%
\small
\Function{HandleCoinCreationRequest}{$R^*$}
\State let $\mathsf{CreateAnonymousCoins}(\pi, C_1, \ldots, C_\ell, B_1 \ldots B_d) = R^*$
\For{$i = 1..\ell$}
    \State ensure $C_i = \cert[R_i]$ is a valid certificate
    \State match $\mathsf{Execute}(\id_i, n_i, \mathsf{Spend}(\val_i, \cm_i, \sigma_i, h_i)) = R_i$
    \State ensure $\cm_i \not \in \{\cm_k\}_{k < i}$
    \State ensure $h_i = \hash(B_1 \ldots B_d)$
\EndFor
\State let $\val = \sum \val_i$
\State verify the ZK-proof $\pi$ on inputs $(\cm_1 \ldots \cm_\ell, \val, B_1 \ldots B_d)$
\State let $s_j = \Call{SignShare}{B_j}$ for each $j = 1..d$
\State \textbf{return} $(s_1, \ldots, s_d)$ \Comment{Return a blinded signature for each output}
\EndFunction
\end{algorithmic}
\end{algorithm}

\fi

\mypara{Redeeming anonymous coins.}
Suppose that a user owns a coin $A = (\id, \cm, \sigma)$.
We define a new account operation $O = \mathsf{SpendAndTransfer}(\id', \sigma, v, r)$
meant to be included in a request $R = \mathsf{Execute}(\id, n, O)$. Following the framework of Section~\ref{sec:accounts}:
\begin{itemize}
    \item $O$ is \emph{safe} iff $\sigma$ is a valid signature for $(\id, \cm)$ with $\cm = \com_{r}(v)$ and $\cm \not\in \spent^\id(\alpha)$.
    \item Upon receiving a valid certificate $C = \cert[R]$, the execution of $O$ consists in adding $\cm$ to $\spent^\id(\alpha)$ and sending a cross-shard request to add the value $v$ to $\bal^{\id'}(\alpha)$ (possibly after creating an empty account $\id'$).
\end{itemize}

The pseudo-code for redeeming operations is presented in Algorithm~\ref{alg:account_operations}.

\mypara{Safety of the protocol.} If $O$ is a transfer operation, we write $\amount(O)$ for the value of the transfer, $\mathsf{source}(O)$ for the main account, $\mathsf{recipient}(O)$ for the recipient account. By extension, we write $\amount(C)$ for the value of a valid confirmation certificate containing such an operation~$O$.

If $A = (\id, \cm, \sigma)$ is valid coin and $\cm = \com_{r}(v)$, we write $\mathsf{id}(A) = \id$, $\textsf{cm}(A) = cm$, and $\amount(A) = v$. We also write $cm \in \spent^\id$ iff there exists a certificate $C = \cert[R]$ with $R = \mathsf{Execute}(\id, n, O)$ and either $O = \mathsf{Spend}(\val, \cm, \sigma, h)$ for some $n$, $\val$, $h$, or $O = \mathsf{SpendAndTransfer}(\id', \sigma, v, r)$ for some $n$, $\id'$.

Under BFT assumption, due to quorum intersection and thanks to the logics related to the spent list in the code of $\mathsf{Spend}$ and $\mathsf{SpendAndTransfer}$, a coin can be spent only once. More precisely, there is one-to-one mapping between certificates $C$ and coins $A$ that justify $\mathsf{cm}(A) \in \spent^\id$ in the definition above.
In what follows, summations over certificates range over all valid certificates for distinct requests or coins.

We define the \emph{spendable value} of an account $\id$ as follows:
\begin{eqnarray*}
\tiny
    \mathsf{spendable}^\id \hspace{-5pt}&=& \hspace{-5pt}\bal^\id(\mathsf{init})
\iffull \;-\; \else \\ &+&\hspace{-10pt}\fi \sum_{\mathsf{recipient}(C) = \id} \hspace{-10pt}\amount(C) - \hspace{-5pt}\sum_{\mathsf{source}(C) = \id}  \hspace{-10pt}\amount(C)
\iffull \;-\; \else \\ &+&\hspace{-10pt}\fi \sum_{\scriptsize\left\{\!\!\!\!\begin{array}{l}\textsf{id}(A)=\id\\ \textsf{cm}(A) \not\in \spent^\id\end{array}\right.}\hspace{-10pt}\amount(A)
\end{eqnarray*}



We verify by inspection of the protocol that the total spendable value over all accounts, that is, $S = \sum_\id \mathsf{spendable}^\id$, never increases during account operations, coin creation, and redeeming of anonymous coins:
\begin{itemize}
    \item Account operations have been studied in Section~\ref{sec:accounts}.
    \item Redeeming coins with a certificate $C$ for $\mathsf{SpendAndTransfer}$ increases the balance of a recipient but burns a coin with corresponding value (i.e. adds it to $\spent^\id$).
    \item Creating coins with a free request $R^* = \mathsf{CreateAnonymousCoins}(\pi, C_1, \ldots, C_\ell, B_1, \ldots, B_d)$ requires withdrawing public amounts and burning the source coins corresponding to $C_1, \ldots, C_\ell$. Importantly, $C_i$ contains a hash commitment of $(B_1, \ldots, B_d)$. Therefore re-reusing the certificate in a coin creation request results in the same coins as the first time, and does not increase~$S$.
\end{itemize}

\mypara{Privacy properties.} The protocol to create anonymous coins guarantees the following privacy properties.
\begin{itemize}
    \item Opacity: Except for the ZK proofs~$\pi$, the coin values under the commitments $\cm^{in}_i$ and $\cm^{out}_j$ are never communicated publicly.
    \item Unlinkability: Assuming that the sender during coin creation is honest, authorities cannot trace back to the origin of an anonymous coin when it is spent.
\end{itemize}
Regarding unlinkability, we note indeed that the receiver information $\id^{out}_j$ and $\cm^{out}_j$ are only communicated to authorities in blinded form. Besides, after unblinding, the threshold signature $\sigma_j$ does not depend on values controlled by authorities, therefore is not susceptible to tainting.

To prevent double spending, the protocol must reveal the identifiers $\id$ of the coins being spent. This means that the sender who initially created the coins linked to $\id$ must be trusted for unlinkability to hold. To mitigate this concern, it is recommended that receivers quickly transfer their new coins anonymously to a secret account so that they can spend them privately later.


\section{Implementation} \label{sec:implementation}

We now sketch our prototype implementation of a multi-core, multi-shard \sysname authority in Rust. Our implementation is based on the existing FastPay codebase\footnote{\url{https://github.com/novifinancial/fastpay}} which already implemented the Byzantine reliable broadcast primitive needed for \sysname. In particular, we were able to re-use modules based on Tokio\footnote{\url{https://tokio.rs}} for asynchronous networking and cryptographic modules based on ed25519-dalek\footnote{\url{https://github.com/dalek-cryptography/ed25519-dalek}} for elliptic-curve-based signatures. For simplicity, data-structures in our \sysname prototype are held in memory rather than persistent storage. Our prototype supports both TCP and UDP for transport. The core of \sysname is idempotent to tolerate retries in case of packet loss. Each authority shard is a separate native process with its own networking and Tokio reactor core.
We are open-sourcing \sysname\footnote{
\ifdefined\cameraReady
\url{https://github.com/novifinancial/fastpay/tree/extensions}
\else
https://www.dropbox.com/s/fdq3l90378r6jjz/zef.zip?dl=0
\fi
} along with any measurements data to enable reproducible results\footnote{
\ifdefined\cameraReady
\url{https://github.com/novifinancial/fastpay/tree/extensions/benchmark_scripts}
\else
https://www.dropbox.com/s/fdq3l90378r6jjz/zef.zip?dl=0
\fi
}.

\paragraph{Cryptographic primitives for anonymous coins}
We have chosen Coconut credentials~\cite{coconut} to implement the blind randomizable threshold-issuance signatures $\sigma^{in}_i$ and $\sigma^{out}_i$ of \Cref{sec:payments}. 
Zero-Knowledge proofs are constructed using standard sigma protocols, made non-interactive through the Fiat-Shamir heuristic~\cite{fiat1986prove}. As a result, our implementation of \sysname assumes the hardness of LRSW~\cite{lysyanskaya1999pseudonym} and XDH~\cite{bls} (required by Coconut), and the existence of random oracles~\cite{fiat1986prove}. 
\Cref{sec:nizk_protocol} presents this protocol in details.
Our implementation of Coconut is inspired from Nym's\footnote{\url{https://github.com/nymtech/coconut}} and uses the curve BLS12-381~\cite{bls12-381} as arithmetic backend.

We have implemented all range proofs using Bulletproofs~\cite{bulletproofs} as they only rely on the discrete logarithm assumption (which is implied by XDH) and do not require a trusted setup. Unfortunately, we couldn't directly use Dalek's implementation of Bulletproofs\footnote{\url{https://github.com/dalek-cryptography/bulletproofs}} as it uses Ristretto~\cite{ristretto} as arithmetic backend. Ristretto is incompatible with Coconut (which requires a pairing-friendly curve). Therefore, we have modified Dalek's implementation to use curve BLS12-381. This required significant effort as the curve operations are deeply baked into the library. Our resulting library is significantly slower than Dalek's for two reasons: operations over BLS12-381 are slower than over Ristretto, and we couldn't take advantage of the parallel formulas in the AVX2 backend present in the original library. We are open-sourcing our Bulletproof implementation over BLS12-381\footnote{
\ifdefined\cameraReady
\url{https://github.com/novifinancial/fastpay/tree/extensions/bulletproofs}
\else
https://www.dropbox.com/s/fdq3l90378r6jjz/zef.zip?dl=0
\fi
}.
\section{Evaluation} \label{sec:evaluation}
We now present our evaluation of the performance of our \sysname prototype based on experiments on Amazon Web Services (AWS).
Our focus was to verify that 
(i) \sysname achieves high throughput even for large committees,
(ii) \sysname has low latency even under high load and within a WAN,
(iii) \sysname scales linearly when adding more shards, and
(iv) \sysname is robust when some parts of the system inevitably crash-fail. Note that evaluating BFT protocols in the presence of Byzantine faults is still an open research question~\cite{twins}.

We deployed a testbed on AWS, using \texttt{m5.8xlarge} instances across 5 different AWS regions: N. Virginia (us-east-1), N. California (us-west-1), Sydney (ap-southeast-2), Stockholm (eu-north-1), and Tokyo (ap-northeast-1). Authorities were distributed across those regions as equally as possible. Each machine provided 10Gbps of bandwidth, 32 virtual CPUs (16 physical core) on a 2.5GHz, Intel Xeon Platinum 8175, 128GB memory, and ran Linux Ubuntu server 20.04. We selected these machines because they provide decent performance and are in the price range of ``commodity servers''.

In the following sections, each measurement in the graphs is the average of 2 independent runs, and the error bars represent one standard deviation\footnote{Error bars are absent when the standard deviation is too small to observe.}. We set one benchmark client per shard (collocated on the same machine) submitting transactions at a fixed rate for a duration of 5 minutes.

\subsection{Regular Transfers}
We benchmarked the performance of \sysname when making a regular transfer as described in \Cref{sec:accounts}.
When referring to \emph{latency} in this section, we mean the time elapsed from when the client submits the request (Step {\small\circled{1}} in \Cref{fig:account_requests}) to when at least one honest authority processes the resulting confirmation certificate (Step {\small\circled{5}} in \Cref{fig:account_requests}). We measured it by tracking sample requests throughout the system.

\begin{figure}[t]
\centering
\includegraphics[width=\columnwidth]{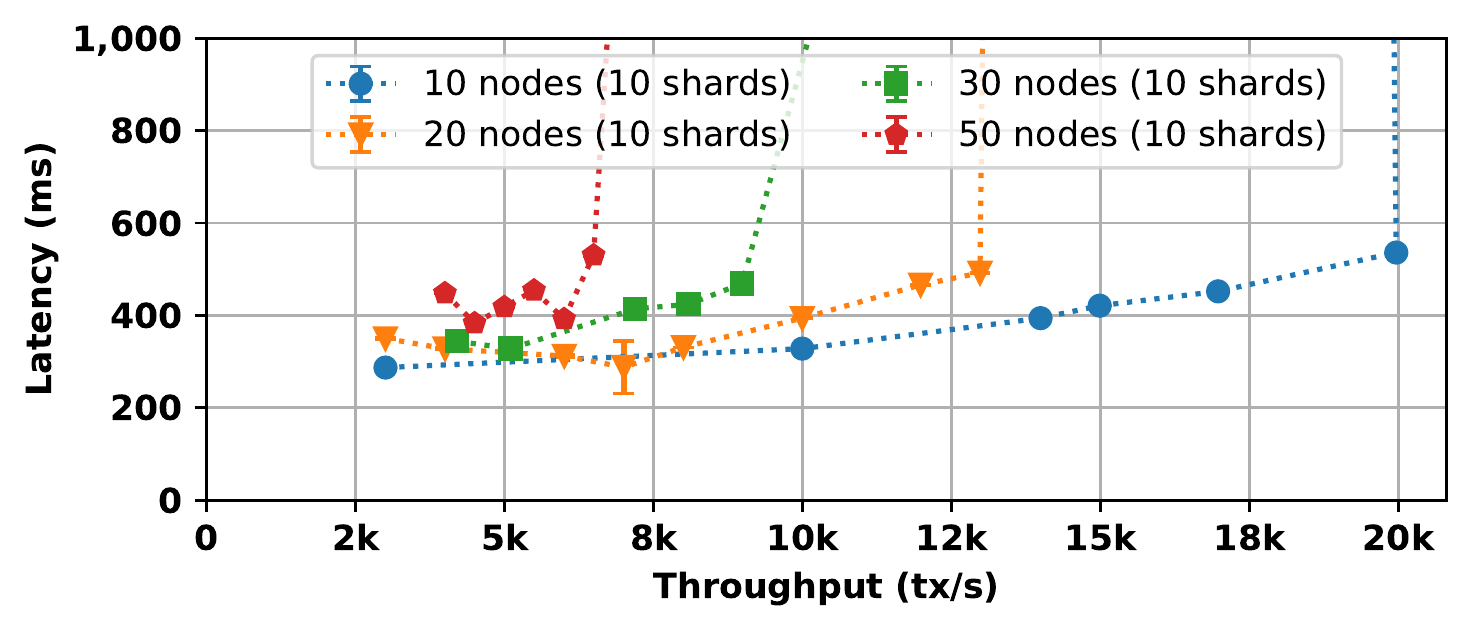}
\caption{
Throughput-latency graph for regular transfers. WAN measurements with 10, 20, 30 authorities; 10 collocated shards per authority. No faulty authorities.
}
\label{fig:common}
\end{figure}

\paragraph{Benchmark in the common case}
\Cref{fig:common} illustrates the latency and throughput of \sysname for varying numbers of authorities. Every authority ran 10 collocated shards (each authority ran thus a single machine). The maximum throughput we observe is 20,000 tx/s for a committee of 10 nodes, and lower (up to 6,000 tx/s) for a larger committee of 50. This highlights the important of sharding to achieve high-throughput. This reduction is due to the need to transfer and check transfer certificates signed by $2f+1$ authorities; increasing the committee size increases the number of signatures to verify since we do not use threshold signatures for regular transfers.

\begin{figure}[t]
\centering
\includegraphics[width=\columnwidth]{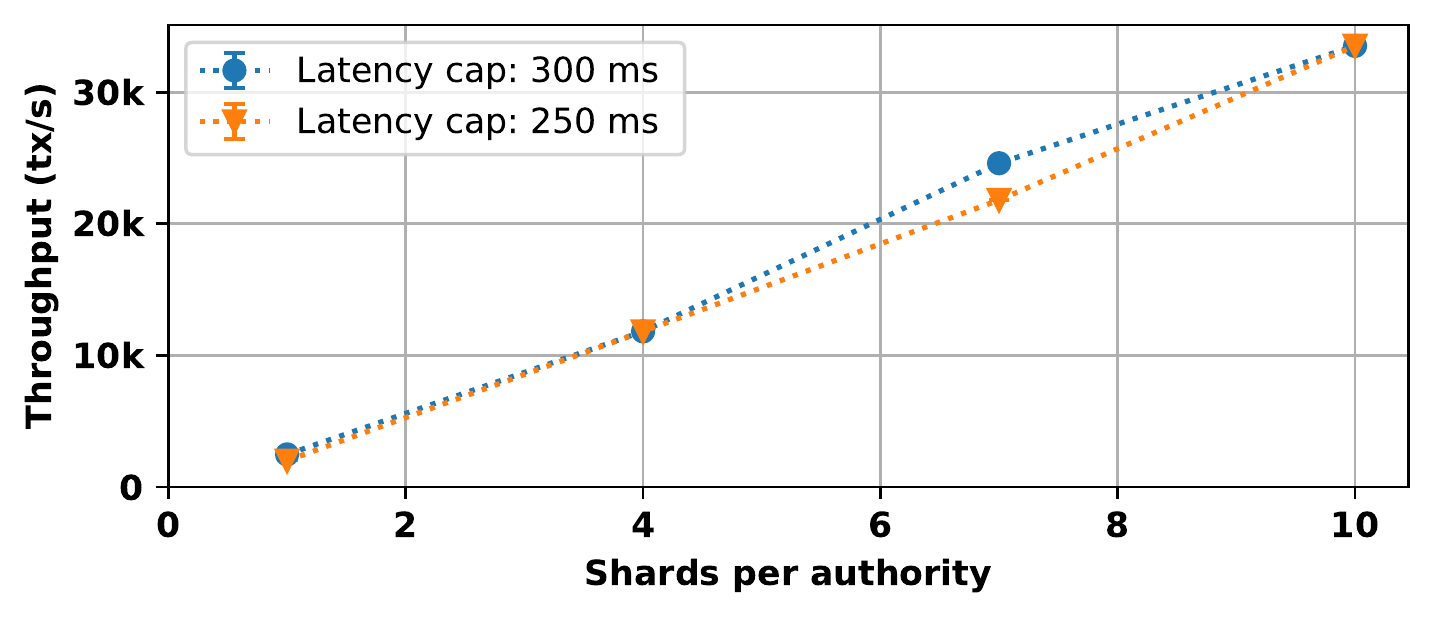}
\caption{
Maximum achievable throughput for regular transfers, keeping the latency under 250ms and 300ms. WAN measurements with 4 authorities; 1 to 10 shards per authority running on separate machines. No faulty authorities.
}
\label{fig:scalability}
\end{figure}

\paragraph{Scalability}
\Cref{fig:scalability} shows the maximum throughput that can be achieved while keeping the latency under 250ms and 300ms. The committee is composed by 4 authorities each running a data-center; each shard runs on a separate machine. \Cref{fig:scalability} clearly supports our scalability claim: the throughput increases linearly with the number of shards, ranging from 2,500 tx/s with 1 shard per authority to 33,000 tx/s with 10 shards per authority.

\begin{figure}[t]
\centering
\includegraphics[width=\columnwidth]{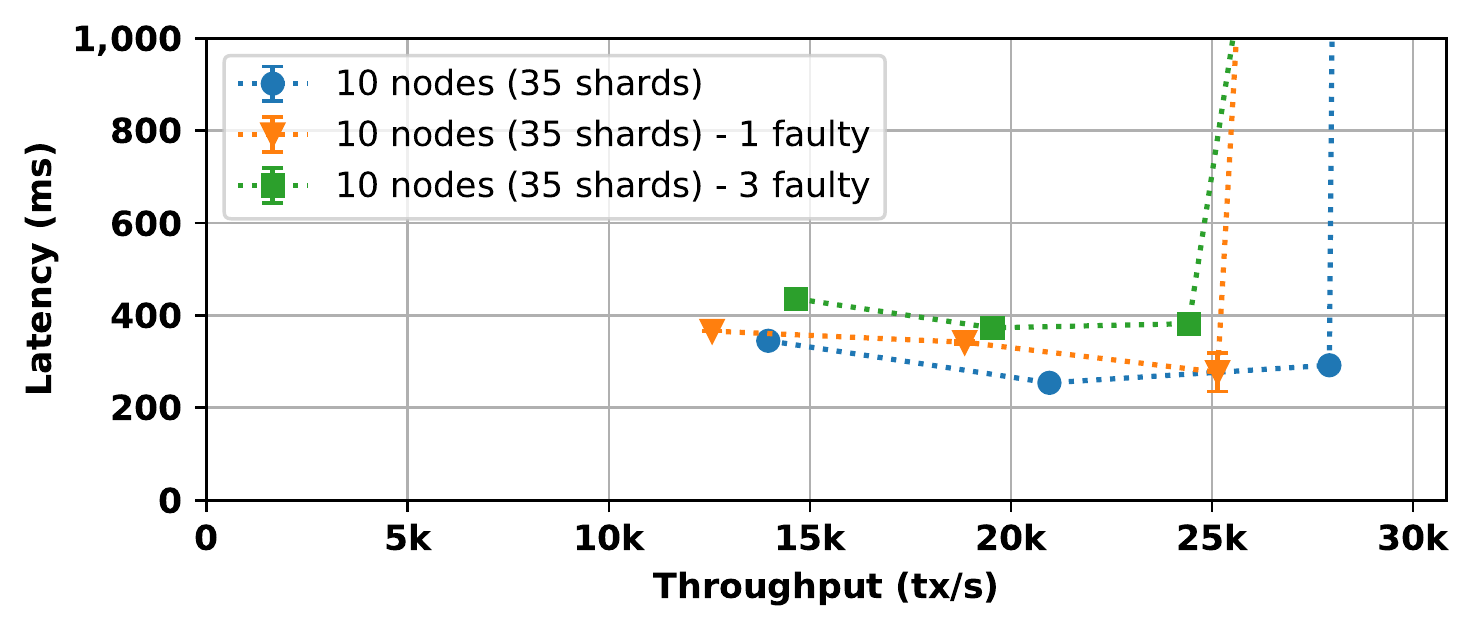}
\caption{
Throughput-latency graph for regular transfers under crash-faults. WAN measurements with 10 authorities; 35 collocated shards per authority; 0, 1, and 3 crash-faults.
}
\label{fig:faults}
\end{figure}

\paragraph{Benchmark under crash-faults}
\Cref{fig:faults} depicts the performance of \sysname when a committee of 10 authorities suffers 1 to 3 crash-faults (the maximum that can be tolerated in this setting). Every authority runs 35 collocated shards (each authority runs thus a single machine).
Contrarily to BFT consensus systems~\cite{DBLP:conf/icdcs/LeeSHKN14}, \sysname maintains a good level of throughput under crash-faults. The underlying reason for the steady performance under crash-faults is that \sysname doesn't rely on a leader to drive the protocol. The small reduction in throughput is due to losing the capacity of faulty authorities. To assemble certificates, the client is now required to wait for all the remaining $2f+1$ authorities and can't simply select the fastest $2f+1$ votes; this accounts for the small increase of latency.
Note that the performance shown in \Cref{fig:faults} are superior to those shown in \Cref{fig:common} because the authorities run more shards. 

\subsection{Anonymous Payments}
We benchmarked the performance of \sysname when spending two opaque coins into two new ones, as described in \Cref{sec:payments}.
When referring to \emph{latency} in this section, we mean the time elapsed from when the client submits the request (Step {\small\circled{2}} in \Cref{fig:anonymous_payments}) to when it assembles the new coins (Step {\small\circled{8}} in \Cref{fig:anonymous_payments}). We measured it by tracking sample requests throughout the system.

\paragraph{Microbenchmarks}
We report on microbenchmarks of the single-CPU core time required to execute the cryptographic operations. \Cref{table:microbenchark} displays the cost of each operation in milliseconds (ms); each measurement is the result of 100 runs on a AWS \texttt{m5.8xlarge} instance. The first 3 rows respectively indicate the time to (i) produce a coin creation request meant to spend two opaque coins into two new ones, (ii) verify that request, and (iii) issue a blinded coin share. The last 3 rows indicate the time to unblind a coin share, verify it, and aggregate 3 coin shares into an output coin.
The dominant CPU cost is on the user when creating a coin request (438.35ms), which involves proving knowledge of each input coins (1 Bulletproof per coin). However, verifying coin requests (142.31ms) is also expensive: it involves verifying the input coins (1 pairing check per input coin) and the output coins request (1 Bulletproof per coin). Issuing a blinded coin share (1 Coconut signature per output coin) is relatively faster (4.90ms). Unblinding (3.37ms), verifying (9.62ms) and aggregating (1.70ms) coin shares take only a few milliseconds.
These results indicate that a single core shard implementation may only settle just over 7 transactions per second---highlighting the importance of sharding to achieve high-throughput.

\begin{table}[t]
\centering
\footnotesize
\begin{tabular}{lrr} \toprule
Measure & Mean (ms) & Std. (ms) \\
\midrule
(User) Generate coin create request & 438.35 & 1.10 \\
(Authority) Verify coin creation request & 142.31 & 0.24 \\
(Authority) Issue a blinded coin share & 4.90 & 0.01 \\
(User) Unblind a coin share & 3.37 & 0.05 \\
(User) Verify a coin share & 9.62 & 0.04 \\
(User) Aggregate 3 coin shares & 1.70 & 0.00 \\
\bottomrule
\end{tabular}
\caption{Microbenchmark of single core CPU costs of anonymous coin operations; average and standard dev. of 100 measurements.}
\label{table:microbenchark}
\end{table}

\begin{figure}[t]
\centering
\includegraphics[width=\columnwidth]{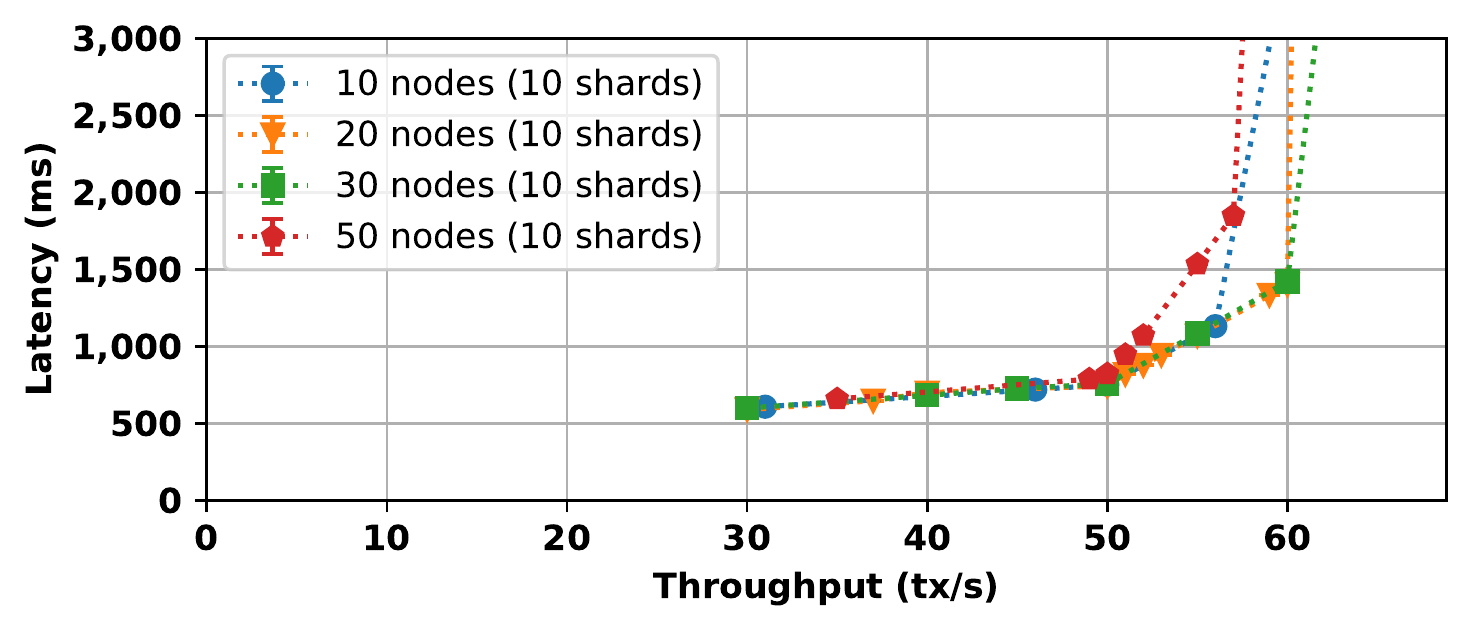}
\caption{
Throughput-latency graph for anonymous coins. WAN measurements with 10, 20, 30 authorities; 10 collocated shards per authority. No faulty authorities.
}
\label{fig:coco-common}
\end{figure}

\paragraph{Benchmark in the common case}
\Cref{fig:coco-common} illustrates the latency and throughput of \sysname for varying numbers of authorities. Every authority runs 10 collocated shards. The performance depicted in \Cref{fig:coco-common} (anonymous payments) are 3 order of magnitude lower than those depicted in \Cref{fig:common} (regular transfers); this is due to the expensive cryptographic operations reported in \Cref{table:microbenchark}.
We observe virtually no difference between runs with 10, 20, 30, or even 50 authorities: \sysname can process about 50 tx/s while keeping latency under 1s in all configurations. This highlights that anonymous payments operations are extremely CPU intensive and that bandwidth is far from being the bottleneck.

\begin{figure}[t]
\centering
\includegraphics[width=\columnwidth]{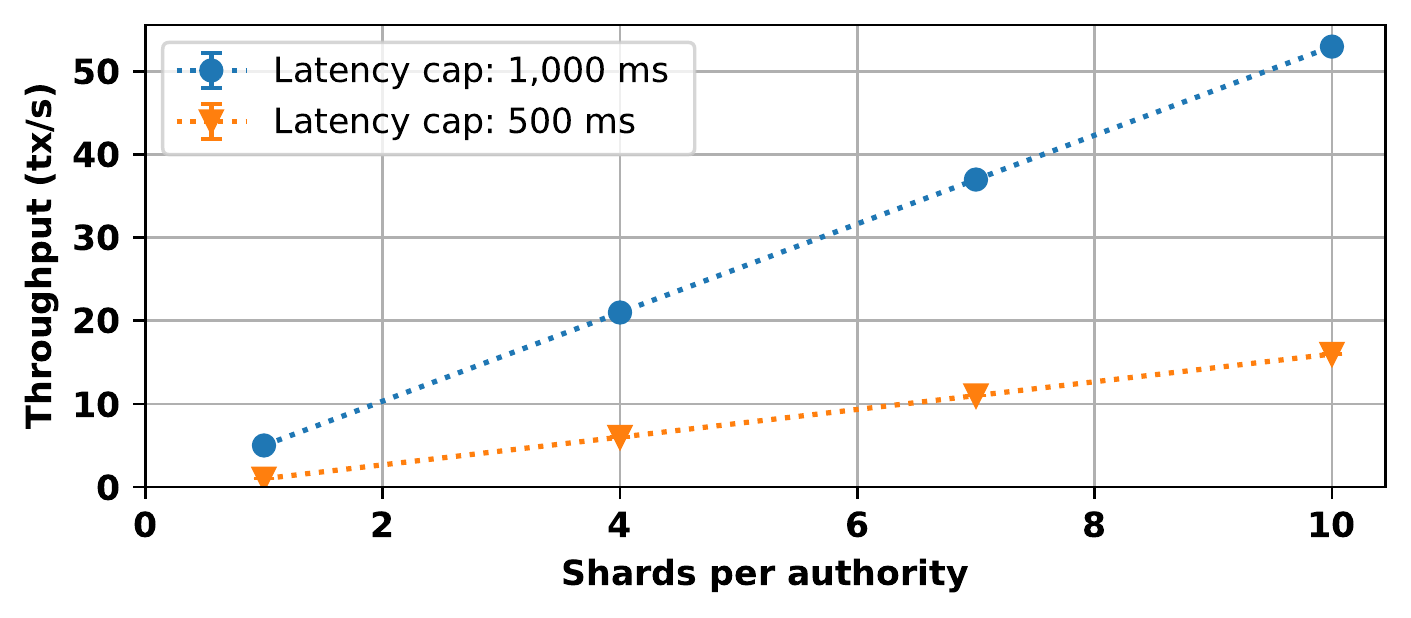}
\caption{
Maximum achievable throughput for anonymous coins while keeping the latency under 500ms and 1s. WAN measurements with 4 authorities; 1 to 10 shards per authority running on separate machines. No faulty authorities.
}
\label{fig:coco-scalability}
\end{figure}

\paragraph{Scalability}
\Cref{fig:coco-scalability} shows the maximum throughput that can be achieved while keeping the latency under 500ms and 1s. The committee was composed by 4 authorities each running a data-center; each shard runs on a separate machine. \Cref{fig:scalability} demonstrates our scalability claim: throughput increases linearly with the number of shards, ranging from 5 tx/s with 1 shard per authority to 55 tx/s with 10 shards per authority (with a latency cap of 1s). 

\begin{figure}[t]
\centering
\includegraphics[width=\columnwidth]{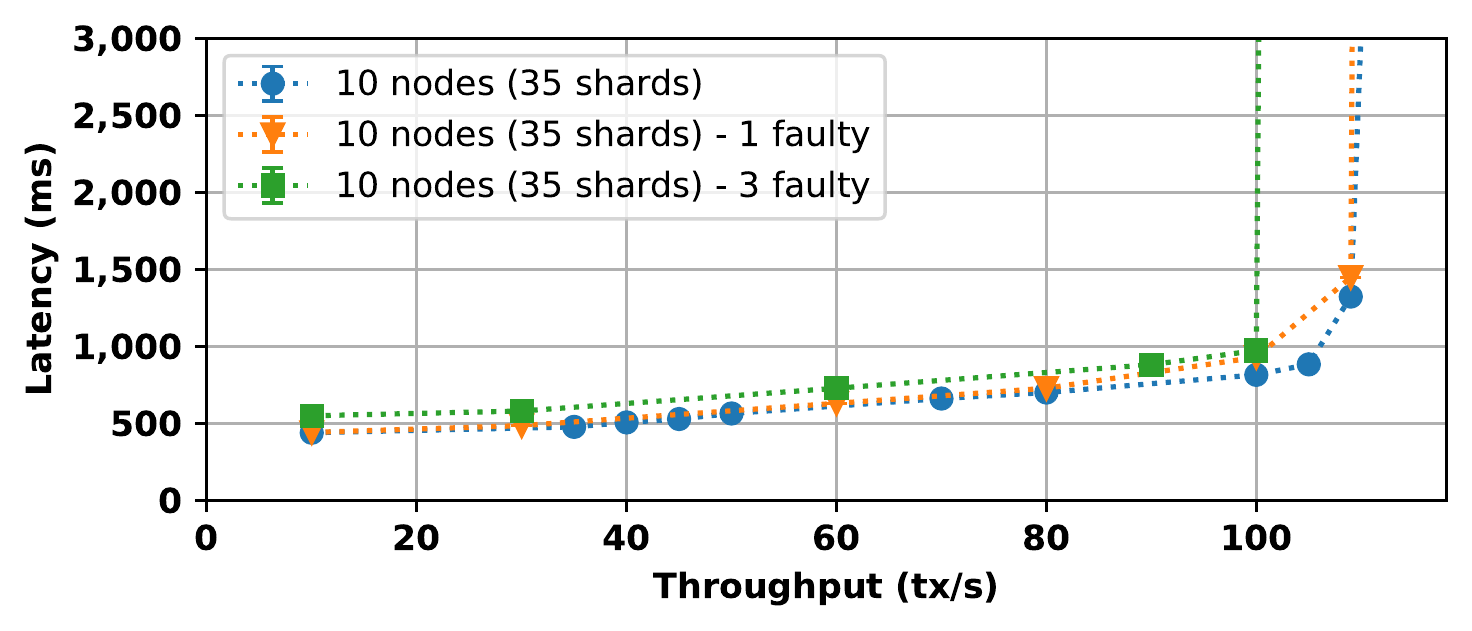}
\caption{
Throughput-latency graph for anonymous coins under crash-faults. WAN measurements with 10 authorities; 35 collocated shards per authority; 0, 1, and 3 crash-faults.
}
\label{fig:coco-faults}
\end{figure}

\paragraph{Benchmark under crash-faults}
\Cref{fig:coco-faults} depicts the performance of \sysname when a committee of 10 authorities suffers 1 to 3 crash-faults. Every authority ran 35 collocated shards (each authority ran thus a single machine).
There is no noticeable throughput drop under crash-faults, and \sysname can process up to 100 tx/s within a second with 0, 1, or 3 faults.
The performance of \sysname shines compared to Zcash~\cite{Ben-SassonCG0MTV14} which is known to process about 27 tx/s with a 1 hour latency~\cite{alfphazero}. Similarly, Monero~\cite{monero} processes about 4 tx/s with a 30 minute latency~\cite{alfphazero}.
%

\section{Conclusion}
\label{sec:conclusion}

\sysname is the first linearly-scalable BFT protocol for anonymous payments with sub-second latency. \sysname follows the FastPay model~\cite{fastpay} by defining authorities as sharded services and by managing singly-owned objects using reliable broadcast rather than consensus. To support anonymous coins without sacrificing storage costs, \sysname introduces a new notion of uniquely-identified, spendable account. Users can bind new anonymous coins to their accounts and spend coins in a privacy-preserving way thanks to state-of-the-art techniques such as the Coconut scheme~\cite{coconut}.

Despite the CPU-intensive cryptographic operations required to preserve opacity and unlinkability of digital coins, our experiments confirm that anonymous payments in \sysname provides unprecedentedly quick confirmation time (sub-second instead of tens of minutes) while supporting arbitrary throughput thanks to the linearly-scalable architecture.

In future work, we wish to explore applications of \sysname beyond payments. To this end, one may consider generalizing account balances using Commutative Replicated Data Types (CmRDTs)~\cite{study-crdts}. Alternatively, one could introduce short-lived instances of a BFT consensus protocol whenever agreements on multi-tenant objects are needed by the system.

\ifdefined\cameraReady
\ifPETS
\section*{Acknowledgments}
The main part of this work was conducted while the authors were at Facebook. This work is mainly supported by Facebook Novi.
\fi
\fi

\iffull \printbibliography \fi
\ifCCS \bibliographystyle{Templates/ACM/ACM-Reference-Format} \bibliography{references} \fi
\ifPETS \bibliographystyle{Templates/ACM/ACM-Reference-Format} \bibliography{references} \fi

\appendix

\ifPETS
\section{Algorithms}
\begin{algorithm}
\caption{Account service (message handlers)}
\label{alg:account_service}
\small
\begin{algorithmic}[1]
\Function{HandleRequest}{$\mathsf{auth}[R]$}
\State let $\mathsf{Execute}(\id, n, O) = R$
\State ensure $\pubk^\id \neq \bot$ \Comment{The account must be active}
\State verify that $\mathsf{auth}[R]$ is valid for $\pubk^\id$ \Comment{Check authentication}
\If {$\pending^\id \neq R$}
    \State ensure $\pending^\id = \bot$
    \State ensure $\nextsequence^\id = n$
    \State ensure \Call{ValidateOperation}{$\id$, $n$, $O$}
    \State $\pending^\id \gets R$ \Comment{Lock the account on $R$}
\EndIf
\State \textbf{return} $\Call{Vote}{R}$ \Comment{Success: return a signature of the request}
\EndFunction
\Statex
\Function{HandleConfirmation}{$C$}
\State verify that $C = \mathsf{cert}[R]$ is valid
\State let $\mathsf{Execute}(\id, n, O) = R$
\State ensure $\pubk^\id \neq \bot$ \Comment{Make sure the account is active}
\If {$\nextsequence^\id = n$}
    \State run \Call{ExecuteOperation}{$\id$, $O$, $C$}
    \State $\nextsequence^\id \gets n + 1$ \Comment{Update sequence number}
    \State $\pending^\id \gets \bot$ \Comment{Make the account available again}
    \State $\confirmed^\id \gets \confirmed^\id :: C$ \Comment{Log certificate}
\EndIf
\EndFunction
\end{algorithmic}
\end{algorithm}

\begin{algorithm}
\caption{Coin creation service}
\label{alg:coin_creation}
\begin{algorithmic}[1]
\algdef{SE}[ASYNC]{Async}{EndAsync}{\textbf{do asynchronously}}{\algorithmicend}%
\algtext*{EndAsync}%
\small
\Function{HandleCoinCreationRequest}{$R^*$}
\State let $\mathsf{CreateAnonymousCoins}(\pi, C_1, \ldots, C_\ell, B_1 \ldots B_d) = R^*$
\For{$i = 1..\ell$}
    \State ensure $C_i = \cert[R_i]$ is a valid certificate
    \State match $\mathsf{Execute}(\id_i, n_i, \mathsf{Spend}(\val_i, \cm_i, \sigma_i, h_i)) = R_i$
    \State ensure $\cm_i \not \in \{\cm_k\}_{k < i}$
    \State ensure $h_i = \hash(B_1 \ldots B_d)$
\EndFor
\State let $\val = \sum \val_i$
\State verify the ZK-proof $\pi$ on inputs $(\cm_1 \ldots \cm_\ell, \val, B_1 \ldots B_d)$
\State let $s_j = \Call{SignShare}{B_j}$ for each $j = 1..d$
\State \textbf{return} $(s_1, \ldots, s_d)$ \Comment{Return a blinded signature for each output}
\EndFunction
\end{algorithmic}
\end{algorithm}

\afterpage{\clearpage}
\fi

\section{Transparent Coins}
\label{sec:transparent_coins}

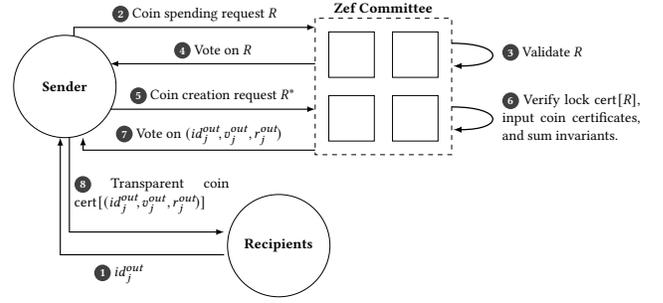
\begin{figure}
  \centering
  \resizeboxcol{\columnwidth}{
    \begin{tikzpicture}
      \node[draw,circle, minimum size=2.25cm] (0,0) {\textbf{Sender}};
      \node[draw,circle, minimum size=2.25cm] at (4.7,-3.5) {\textbf{Recipients}};

      \tikzmath{\sidelen=3;}
      \tikzmath{\s=\sidelen/10;\inlen=\sidelen/3;}
      \begin{scope}[shift={(5.5,1.5)}]
        \draw[dashed] (0,0) rectangle (\sidelen,-1*\sidelen);
        \node[above] at (\sidelen/2,0) {\textbf{\sysname Committee}};

        \draw (\s,{-1*\s} ) rectangle ({\s+\inlen},{-1*(\s+\inlen)} );
        \draw ({\sidelen - \s - \inlen},{-1*\s} ) rectangle ({\sidelen-\s},{-1*(\s+\inlen)} );
        \draw (\s,{-1*(\sidelen - \s - \inlen)}) rectangle ({\s+\inlen},{-1*(\sidelen-\s)});
        \draw ({\sidelen - \s - \inlen},{-1*(\sidelen - \s - \inlen)}) rectangle ({\sidelen-\s},{-1*(\sidelen-\s)});
      \end{scope}

      \draw[solid,-latex, line width=0.25mm] (0.2,1.12) -- (0.2,1.3) -- (5.5,1.3) node[midway,above]{\circled{2} Coin spending request $R$};
      \draw[solid,-latex, line width=0.25mm] (5.5,0.5) -- (1,0.5) node[midway,above]{\circled{4} Vote on $R$};
      \draw[solid,-latex, line width=0.25mm] (1,-0.5) -- (5.5,-0.5)  node[midway,above]{\circled{5} Coin creation request $R^*$};
      \draw[solid,-latex, line width=0.25mm] (5.5,-1.4) -- (0.4,-1.4) node[midway,above]{\circled{7} Vote on $(\id^{out}_j\!, v^{out}_j, r^{out}_j\!)$} -- (0.4,-1.12);

      \begin{scope}[shift={(5.5,1.5)}]
        \draw[thick, -latex, shorten >=1pt] (\sidelen,{-1*(\s+\inlen/4)}) to [out=0,in=0,loop,looseness=6] (\sidelen,{-1*(\s+3*\inlen/4)}) node[right,shift={(1,\s)}] {\circled{3} Validate $R$};
        \draw[thick, -latex, shorten >=1pt] (\sidelen,{-1*(\sidelen - (\s+3*\inlen/4))}) to [out=0,in=0,loop,looseness=6] (\sidelen,{-1*(\sidelen - (\s+\inlen/4))}) node[right,shift={(1,\s)}] {\parbox{3cm}{\circled{6} Verify lock $\cert[R]$, input coin certificates, and sum invariants.}};
      \end{scope}

      \draw[solid,-latex, line width=0.25mm] (0.1,-1.125) -- (0.1,-3.2) node[midway,right,shift={(0,-0.225)}]{\parbox{3.4cm}{\circled{8} Transparent coin $\cert[(\id^{out}_j\!, v^{out}_j, r^{out}_j\!)]$}} -- (3.5,-3.2) ;

      \draw[solid,-latex, line width=0.25mm] (3.5,-3.7) -- (-0.1,-3.7) node[midway,below,shift={(-0.5,-0.1)}]{\circled{1} $\id^{out}_j$} -- (-0.1,-1.125) ;

    \end{tikzpicture}
  }
  \caption{A payment with transparent coins}
  \label{fig:transparent_payments}
\end{figure}

For comparison purposes, we sketch a simplified version of anonymous coins (Section~\ref{sec:payments}) without opacity and unlinkability.
At a high level, the protocol is similar to anonymous coins in terms of communication (Figure~\ref{fig:transparent_payments}). Due to the absence of blinding and random commitments, communication channels and validators must be trusted for the privacy of every coin operation.

\mypara{Transparent coins.}

A \emph{transparent coin} is a certificate $T = \cert[S]$ on a triplet $S = (\id, v, r)$ where $\id$ is the identifier of an account, $v \in [0, v_\textsf{max}]$, and $r$ is some random seed value.
Seed values~$r$ are used to distinguish coins of the same value attached to the same $\id$.

To spend a transparent coin $T$, a client must possess the authentication key controlling $\id$.
Importantly, authorities do not need to store $T$ themselves---although they will observe such certificates occasionally in clear.

\mypara{New account operation.}
Similar to Section~\ref{sec:payments}, we assume a new account operation $O = \mathsf{Spend}(\val, T, h)$ meant to prepare the creation of new coins associated to $h$, by consuming a coin $T$ and by withdrawing an amount $\val$ publicly.
Consider an operation $O = \mathsf{Spend}(\val, T, h)$ included in a request $R = \mathsf{Execute}(\id, n, O)$.
\begin{itemize}
    \item $O$ is \emph{safe} iff $0 \leq \val \leq \bal^\id(\alpha)$, $T = \cert[S]$ is a valid certificate for $S = (\id, v, r)$, and $r \not\in \spent^\id(\alpha)$.
    \item The execution of $O$ consists in adding $r$ to $\spent^\id(\alpha)$ and subtracting $\val$ from $\bal^\id(\alpha)$.
\end{itemize}

\mypara{Transparent coin payment protocol.}

Suppose that a user owns $\ell$ mutually distinct transparent coins $T^{in}_i = \cert[S^{in}_i]$ where $S^{in}_i = (\id^{in}_i, v^{in}_i, r^{in}_i)$ ($1 \leq i \leq \ell$). Let $\val_i \geq 0$ be a value that the user wishes to withdraw publicly from the account $\id^{in}_i$.
Similar to Section~\ref{sec:payments}, we require certificates $T^{in}_i$ to be distinct but not the identifiers $\id^{in}_i$.
We define the total input value of the transfer as $v = \sum_i v^{in}_i + \sum_i \val_i$.

To spend the coins into $d$ new coins with values $v^{out}_j$ ($1 \leq j \leq d$) such that $\sum_j v^{out}_j = v$, the sender requests an identifier $\id^{out}_j$ from each recipient, then proceeds as follows:
\begin{enumerate}
\item For every $1 \leq j \leq d$, sample randomness $r^{out}_j$. Let $S^{out}_j = (\id^{out}_j, v^{out}_j, r^{out}_j)$.

\item For every input $i$, the sender obtains a certificate $C_i = \cert[R_i]$ and executes a request $R_i = \mathsf{Execute}(\id^{in}_i, n_i, O_i)$ where $O_i = \mathsf{Spend}(\val^{in}_i, T^{in}_i, \hash(S^{out}_1, \ldots, S^{out}_d))$, $n_i$ is the next available sequence number for the account~$\id^{in}_i$.

\item Next, the sender broadcasts a free request $R^* = \mathsf{CreateTransparentCoins}(C_1, \ldots, C_\ell, S^{out}_1, \ldots, S^{out}_d)$ and waits for a quorum of responses.

\item Upon receiving a free request of the form $R^* = \mathsf{CreateTransparentCoins}(C_1, \ldots, C_\ell, S_1, \ldots, S_d)$ where $S_j = (\id^{out}_j, v^{out}_j, r^{out}_j)$, each authority $\alpha$ verifies the following:
\begin{itemize}
    \item $C_i = \cert[R_i]$ is a valid certificate for a request of the form $R_i = \mathsf{Execute}(\id^{in}_i, n_i, O_i)$ where $O_i = \mathsf{Spend}(\val^{in}_i, T_i, h_i)$,
    \item The certificates $T_i$ are mutually distinct.
    \item $\sum_i v^{in}_i \,+\, \sum_i \val_i = \sum_j v^{out}_j$.
\end{itemize}
The authority then responds with one signature for each $S^{out}_j$.

\item For every $j$, the sender finally combines a quorum of signatures on $S^{out}_j$ into a new coin $T^{out}_j$.

\item The $j^{th}$ recipient receives $T^{out}_j = \cert[(\id^{out}_j, v^{out}_j, r^{out}_j)]$. She verifies that the values and the identifiers are as expected, that the random seeds $r^{out}_j$ are mutually distinct, and that the certificates $T^{out}_j$ are valid.
\end{enumerate}

\mypara{Redeeming transparent coins.}

Suppose that a user owns a transparent coin $T$ linked to the account $\id$.
We define a new account operation $O = \mathsf{SpendAndTransfer}(\id', T)$
meant to be included in a request $R = \mathsf{Execute}(\id, n, O)$. Following the framework of Section~\ref{sec:accounts}:
\begin{itemize}
    \item $O$ is \emph{safe} iff $T = \cert[S]$ is a valid certificate for $S = (\id, v, r)$ and $r \not\in \spent^\id(\alpha)$.
    \item Upon receiving a valid certificate $C = \cert[R]$, the execution of $O$ consists in adding $r$ to $\spent^\id(\alpha)$, then sending a cross-shard request to add the value $v$ to $\bal^{\id'}(\alpha)$ (possibly after creating an empty account $\id'$).
\end{itemize}

\newcommand{\bbG}{\mathbb{G}}
\newcommand{\ProveCred}{\textsf{ProveCred}}
\newcommand{\VerifyCred}{\textsf{VerifyCred}}
\newcommand{\Randomize}{\textsf{Randomize}}
\newcommand{\PrepareBlindSign}{\textsf{PrepareBlindSign}}
\newcommand{\BlindSign}{\textsf{BlindSign}}
\newcommand{\AggAndUnblind}{\textsf{AggAndUnblind}}
\newcommand{\Unblind}{\textsf{Unblind}}

\newcommand{\AggCred}{\textsf{AggCred}}

\newcommand{\qquadand}{\qquad{\rm and}\qquad}
\newcommand{\quadand}{\quad{\rm and}\quad}
\newcommand{\NIZK}{{\rm NIZK}}

\newcommand{\tabifnotfull}{\iffull\else\hspace*{1em}\fi}

\newcommand\alg[2]{\ding{118}\xspace \textnormal{\textsf{#1}}\xspace$\bm{\rightarrow}$\xspace(#2):\xspace}

\newcommand{\twocolalgnoding}[2]{\iffull\alg{#1}{#2}\else\textnormal{\textsf{#1}}\xspace$\bm{\rightarrow}$\xspace(#2):\xspace\fi}

\section{NIZK Protocol}

\label{sec:nizk_protocol}
In this section, we show one possible efficient instantiation of the anonymous payment protocol from Section~\ref{sec:payments} by opening up the cryptographic primitives used. Our protocol here makes use of the Coconut threshold credential scheme~\cite{coconut}, which is based on the work of Pointcheval and Sanders~\cite{pointcheval2016sig}. Informally, Coconut allows users to obtain credentials on messages with private attributes in a distributed setting using a threshold $t$ out of $n$ authorities.

\renewcommand{\vec}{\bar}

\subsection{Coconut++}
\label{subsec:coconut++}
We start by giving an overview of a suitable variant of the Coconut scheme, nicknamed Coconut++. This variant of Coconut is formally proven secure by Rial and Piotrowska~\cite{rial2022security}.
At a high level, Coconut allows a user to obtain, from a threshold number of authorities, an anonymous credential on a private attribute $m$ showing that it satisfies some application-specific predicate $\phi(m) = 1$. Later, the user can anonymously prove the validity of this credential to any entity in possession of the verification key. While the standard Coconut scheme works for a single attribute,~\cite{coconut} also includes an extension that allows for credentials on a list of $q$ integer-valued attributes $\vec m = (m_1, \dots, m_q)$.

Below, we use the notation $\vec X = (X_1, \dots, X_q)$ for any list of $q$ variables $X_i$ ($1 \leq i \leq q$). The scheme Coconut++ consists of the following algorithms:

\begin{description}[leftmargin=1em, labelindent=0em]
\setlength\itemsep{0.5em}

\item[\alg{Setup($1^\lambda$)}{$pp$}]
Choose groups $(\bbG_1, \bbG_2, \bbG_T)$ of order $p$ (a $\secparam$-bit prime) with a bilinear map $e:\bbG_1 \times \bbG_2 \to \bbG_T$. Let $H: \bbG_1 \to \bbG_1$ be a secure hash function. Let $g_1, h_1, \ldots, h_q$ be generators of $\bbG_1$ and let $g_2$ be a generator of $\bbG_2$. The system parameters are given as $pp = (\bbG_1, \bbG_2, \bbG_3, p, e, H, g_1, g_2, \vec h)$. Parameters are implicit in the remaining descriptions.

\item[\alg{KeyGen($t, n$)}{$\sk, \vk$}]
Pick $q+1$ polynomials $u, w_1, \ldots, w_q$ each of degree $t-1$ with coefficients in $\F_p$ and set $\sk = (x, \vec y) = \left(u(0), w_1(0), \ldots, w_q(0) \right)$. Publish the verification key $\vk = (\vec \gamma, \alpha, \vec \beta) = (g_1^{y_1}, \ldots, g_1^{y_q}, g_2^x, g_2^{y_1}, \ldots, g_2^{y_q})$. Also issue to each authority $j \in \{1, \dots, n\}$, the secret key $\sk_j = (x_j, \vec y_j) = (u(j), w_1(j), \ldots, w_q(j))$ and publish the corresponding verification key $\vk_j = (\vec \gamma_j, \alpha_j, \vec \beta_j) = (g_1^{y_{j,1}}, \ldots, g_1^{y_{j,q}}, g_2^{x_j}, g_2^{y_{j,1}}, \ldots, g_2^{y_{j,q}})$.

\item[\alg{PrepareBlindSign($\vec m, \phi$)}{$\vec r, \Lambda$}]
Pick a random $o \in \F_p$. Compute the commitment $c_{\vec m}$ and group element $h$ as
\[
    c_{\vec m} = g_1^o \prod^q_{i = 1} h_i^{m_i}  \qquadand h = H(c_{\vec m})
\]

\noindent For all $i = 1\ldots q$, pick a random $r_i \in \F_p$ and compute the blinded value ${c}_i$ as follows:

\[
    c_i = h^{m_i} g_1^{r_i}
\]

\noindent Output ($\vec r, \Lambda$) where $\Lambda = (c_{\vec m}, \vec c, \pi_s)$ where $\pi_s$ is defined as:

\iffull
\[
    \pi_{s} = {\rm NIZK}\{(\vec m, o, \vec r): \forall i, c_i = h^{m_i} g_1^{r_i} \;\land\; c_{\vec m} = g_1^o \prod^q_{i = 1} h_i^{m_i} \;\land\;  \phi(\vec m)=1\}
\]
\else
\begin{align*}
    \pi_{s} = {\rm NIZK}\{(\vec m, o, \vec r):\;\; &\forall i, c_i = h^{m_i} g_1^{r_i} \;\land\; c_{\vec m} = g_1^o \prod^q_{i = 1} h_i^{m_i} \\ & \land\;  \phi(\vec m)=1\}
\end{align*}
\fi

\item[\alg{BlindSign($\sk_j, \Lambda, \phi$)}{$\widetilde{\sigma}_j$}] The authority $j$ parses $\Lambda = (c_{\vec m}, \vec c, \pi_{s})$, and $\sk_j = (x_j, \vec y_j)$. Recompute $h = H(c_{\vec m})$. Verify the proof $\pi_{s}$ using $\vec c, c_{\vec m}$ and $\phi$; if the proof is valid, compute $\widetilde{s}_j = h^{x_j} \prod^q_{i=1} c_i^{y_{j,i}}$ and output $\widetilde{\sigma}_j = (h, \widetilde{s}_j)$; otherwise output $\perp$.

\item[\alg{Unblind($\widetilde{\sigma}_j, \vec r, \vec \gamma$)}{$\sigma_j$}] Parse $\widetilde{\sigma}_j=(h, \widetilde{s_j})$, let $s_j = \widetilde{s_j} \prod^q_{i=1} \gamma_i^{-r_i}$, and output $\sigma_j = (h, s_j)$.

This results in $\sigma_j = (h, s_j)$ where $s_j = h^{x_j} \prod^q_{i=1} c_i^{y_{j,i}} \prod^q_{i=1} \gamma_i^{-r_i} = h^{x_j + \sum^q_{i=1} y_{j,i}\,m_i}$.

This is similar to a Waters signature~\cite{waters2005efficient} related to the public key of each authority. Verification of partial coins is used in the implementation of \sysname for clients to validate a quorum of answers received in parallel from authorities and discard erroneous values before running the aggregation step.

\item[\alg{AggCred($\{\sigma_j\}_{j \in J}$)}{$\sigma$}] Return $\bot$ if $\len{J} \neq t$. Parse each $\sigma_j$ as $(h, s_j)$. Output $\sigma = (h, \prod_{j\in J} s_j^{\ell_j})$, where each $\ell_j$ is the Lagrange coefficient given by:
\begin{equation}\nonumber
\ell_j = \left[\prod_{k\in I\setminus\{j\}} (0-k)\right] \left[\prod_{k\in I\setminus\{j\}} (j-k)\right]^{-1} \;{\rm mod}\; p
\end{equation}

This computation results in a value $\sigma = (h, h^{x + \sum^q_{i=1} y_i\,m_i})$ that does not depend on the set of authorities~$J$.

\item[\alg{ProveCred($\vk, \vec m, \sigma, \phi'$)}{$\Theta, \phi'$}]
Parse $\sigma=(h,s)$ and $\vk=(\vec \gamma, \alpha, \vec \beta)$. Pick at random $r, r' \in \mathbb{F}_p^2$, set $h' = h^{r'}$, $s' = s^{r'} (h')^r$, and
$\sigma' = (h',s')$. Build $\kappa = \alpha \, g_2^r \prod^q_{i=1} \beta_i^{m_i}$. Then, output $(\Theta, \phi')$, where $\Theta = (\kappa, \sigma',\pi_v)$ and $\phi'$ is an application-specific predicate satisfied by $\vec m$, and $\pi_v$ is:
\begin{equation}\nonumber
    \pi_v={\rm NIZK}\{(\vec m, r): \kappa=\alpha \, g_2^r \prod^q_{i=1} \beta_i^{m_i} \ \land \  \phi'(\vec m) = 1\}
\end{equation}

\item[\alg{VerifyCred($\vk, \Theta, \phi'$)}{$true/false$}] Parse $\Theta = (\kappa, \sigma',\pi_v)$ and $\sigma'=(h',s')$; verify $\pi_v$ using $\vk$ and $\phi'$. Output $true$ if the proof verifies, $h'\neq1$ and the bilinear evaluation $e(h',\kappa)=e(s', g_2)$ holds; otherwise output $false$.

The bilinear evaluation is justified by the following equations:
\begin{align*}
e(h', \kappa) &= e(h^{r'}, \alpha \, g_2^r \prod^q_{i=1} \beta_i^{m_i}) = e(h^{r'}, g_2^{x + r + \sum_i y_i\,m_i}) \\
e(s', g_2) &= e(s^{r'} (h')^r, g_2) = e(h^{r'(x + \sum_i y_i\,m_i)}\,h^{r r'}, g_2) \\
\end{align*}
\end{description}


\subsection{Anonymous Transfer Protocol}
We now instantiate the anonymous transfer protocol from Section~\ref{sec:payments} using the Coconut scheme with three attributes $\vec m = (k, q, v)$ consisting of a key~$k$, a random seed $q$, and a private coin value~$v$.
From the point of view of its owner, an opaque coin is defined as $A = (\id, x, q, v, \sigma)$ where $\id$ is the linked account, $x$ is an unique index within the same account $\id$, $q$ is a secret random seed, $v$ is the value of the coin, and $\sigma$ denotes the Coconut credential for $k = \hash(\id :: [x])$, $q$ and $v$. When a new opaque coin is created, the three attributes are hidden to authorities. The account $\id$ and the index $x$ of a coin are revealed when it is spent to verify coin ownership and prevent double-spending of coins within the same account. We use the third attribute $q$ to guarantee the privacy of the value $v$ even after $k$ is revealed\footnote{
As noted in the original Coconut paper~\cite{coconut}, if a credential contains a single attribute $m$ of low entropy (such as a coin value), the verifier can run multiple times the verification algorithm making educated guesses on the value of $m$ and effectively recover its value through brute-force.}.

Suppose that a sender owns $\ell$ input coins $A^{in}_i = (\id^{in}_i, x^{in}_i, q^{in}_i, v^{in}_i, \sigma^{in}_i)$ ($1 \leq i \leq \ell$) and wishes to create $d$ output coins of the form $(\id^{out}_j, x^{out}_j, q^{out}_j, v^{out}_j, \sigma^{out}_j)$ ($1 \leq j \leq d$). Let $\val^{in}_i \geq 0$ denotes a public value to withdraw from the account~$\id^{in}_i$ as in Section~\ref{sec:payments}. The sender must ensure that $\sum_{i} v^{in}_i + \sum_{i}\val^{in}_i = \sum_{j} v^{out}_j$ and that the coin indices $(\id^{out}_j, x^{out}_j)$ are mutually distinct.

\paragraph{Using Coconut for opaque coin transfers.}
We present an overview of the changes to the anonymous transfer protocol from Section~\ref{sec:payments} to implement opaques coins.

Recall that the sender must first construct blinded descriptions of the desired output coins. These descriptions are meant to be incorporated into a hash commitment $h$ in the spending certificates $C_i$ for input coins.
To do so, the sender proceeds as follows. Define $\phi'$ is a predicate satisfied by the input and output coin values and defined as follows: $\phi'(\vec v^{in}, \vec v^{out}) = \true$ iff
\[
\sum_{i}^l v^{in}_i + \sum_{i}^l \val^{in}_i = \sum_{j}^d v^{out}_j
\quad\wedge\quad v^{out}_i \in [0, v_\textsf{max}]
\]
The predicate $\phi'$ binds the NIZKs associated with all $\ProveCred$ proofs for the input coins and all $\PrepareBlindSign$ proofs for the output coins. It also shows that the value on both sides of the transfer is consistent.

For every $1 \leq i \leq \ell$, considering $k^{in}_i = \hash(\id^{in}_i :: [x^{in}_i])$ as public parameters, the sender calls
\[
    \Theta_i \gets \ProveCred(\vk, (q^{in}_i,v^{in}_i),\sigma^{in}_i, \phi')
\]
\noindent Then, for every $1 \leq j \leq d$, she calls
\[
    ((\rk_j, \rq_j, \rv_j),\Lambda_j) \gets \PrepareBlindSign(k^{out}_j, q^{out}_j, v^{out}_j, \phi')
\]

Define $P = (\Theta_1, \ldots, \Theta_\ell, \Lambda_1, \ldots, \Lambda_j, \phi')$ and $h = \hash(P)$. The sender obtains $C_i = \cert[R_i]$ by broadcasting a request $R_i = \mathsf{Execute}(\id_i, n_i, \mathsf{Spend}(\val^{in}_i, x^{in}_i, h))$ for some suitable sequence number $n_i$.
The operation $\mathsf{Spend}$ behaves as the one described in Section~\ref{sec:payments} except that (i)~the attribute $x$ plays the role of $cm$ w.r.t. the spent list $\spent^\id(\alpha)$; and (ii)~for simplicity, we differ the validation of each input coin credential (formerly the signature~$\sigma$ in $O$) to the next step.

Next, the sender submits a request $R^* = \mathsf{CreateAnonymousCoins}(C_1, \ldots, C_\ell, P)$. On receiving $R^*$ from the sender, an authority $\chi$ now verifies the proofs $\Theta_i$ and $\Lambda_j$ and the predicate $\phi'$ by running $\VerifyCred(\vk,  \Theta_i, \phi')$ for each~$i$ and $\widetilde{\sigma}^{out}_j = \BlindSign(\sk_\chi, \Lambda_j, \phi')$ for each $j$. If the proofs are valid, it returns $ \widetilde{\vec \sigma}^{out}$ to the sender.

After collecting $t$ such responses, the sender can now run $\Unblind$ and $\AggCred$ to obtain a valid credential on each created output coin. Finally, to complete the transfer, it can send the coin $(\id_j^{out}, x_j^{out}, q_j^{out}, v_j^{out}, \sigma_j^{out})$ to the $j^\thtext$ recipient.

\paragraph{Opaque coin construction.}
We present the cryptographic primitives used by the opaque coins transfer protocol. The \textsf{Setup} and \textsf{KeyGen} algorithms are exactly the same as Coconut.

\begin{description}[leftmargin=1em, labelindent=0em]
\setlength\itemsep{0.5em}

\iffull
\item[\alg{CoinRequest($\vk, \vec \sigma^{in}, \vec q^{in}, \vec v^{in}, \vec k^{out}, \vec q^{out}, \vec v^{out}, \val^{in}_1, \dots, \val^{in}_\ell$)}{$(\vec \rk, \vec \rq, \vec \rv), \Gamma$}]
\else
\item[\ding{118}]\xspace\twocolalgnoding{CoinRequest($\vk, \vec \sigma^{in}, \vec q^{in}, \vec v^{in}, \vec k^{out}, \vec q^{out}, \vec v^{out}, \val^{in}_1, \dots, \val^{in}_\ell$)}{$(\vec \rk, \vec \rq, \vec \rv), \Gamma$}
\fi

Parse $\vk=(\gamma_0, \gamma_1, \gamma_2, \alpha, \beta_0, \beta_1, \beta_2)$. For every input coin $\sigma^{in}_i$ ($1 \leq i \leq \ell$), parse $\sigma^{in}_i=(h_i, s_i)$, pick at random $\rh_i, \rs_i \in \mathbb{F}_p^2$, and compute
\[
    h'_i = h_i^{\rh_{i}}
    \qquadand s'_i = s_i^{\rh_{i}} (h'_i)^{\rs_{i}}
\]

\noindent Then set $\sigma'^{in}_i = (h'_i, s'_i)$ and build:
\[
    \kappa_i = \alpha \, g_2^{\rs_{i}} \beta_1^{q_i^{in}} \beta_2^{v_i^{in}}
\]

\noindent For every output coin $j$ ($1 \leq j \leq d$), pick a random $o_j \in \F_p$, and compute the commitments $\cm_{j}$ and the group elements $\hat{h}_j$ as
\[
    \cm_{j} = g_1^{o_j} h_0^{k_j^{out}} h_1^{q_j^{out}} h_2^{v_j^{out}}  \qquadand \hat{h}_j = H(\cm_{j})
\]

\noindent For all $1 \leq j \leq d$, pick a random $(\rk_{j}, \rq_{j}, \rv_{j}) \in \F_p^3$ and compute the commitments $(\ck_{j}, \cq_{j}, \cv_{j})$ as follows:
\[
    \tabifnotfull \ck_{j} = \hat{h}_j^{k_j^{out}} g_1^{r_{k,j}}
    \quadand \cq_{j} = \hat{h}_j^{q_j^{out}} g_1^{\rq_{j}}
    \quadand \cv_{j} = \hat{h}_j^{v_j^{out}} g_1^{\rv_{j}}
\]

\noindent Output ($(\vec \rk, \vec \rq, \vec \rv), \Gamma$) where $\Gamma = (\vec \sigma'^{in}, \vec \kappa, \vec \cm, \vec \ck,  \vec \cq, \vec \cv, \pi_r)$ where $\pi_r$ is defined as:

\begin{align*}
    &\pi_{r} = {\rm NIZK}\{
    (\vec q^{in}, \vec v^{in}, \vec k^{out}, \vec q^{out}, \vec v^{out}, \vec \rs, \vec o, \vec r_k, \vec r_q, \vec r_v): \\
    &\qquad \forall i, \kappa_i = \alpha \, g_2^{\rs_{i}} \beta_1^{q_i^{in}} \beta_2^{v_i^{in}} \\
    &\qquad \land\quad \forall j, \cm_{j} = g_1^{o_j} h_0^{k_j^{out}} h_1^{q_j^{out}} h_2^{v_j^{out}} \\
    &\qquad \land\quad \forall j, \ck_{j} = \hat{h}_j^{k_j^{out}} g_1^{\rk_{j}}
    \iffull\else\\&\quad\fi \quad\land\quad \forall j, \cq_{j} = \hat{h}_j^{q_j^{out}} g_1^{\rq_{j}}
    \iffull\else\\&\quad\fi \quad\land\quad \forall \cv_{j} = \hat{h}_j^{v_j^{out}} g_1^{\rv_{j}} \\
    &\qquad \land\quad \sum_{i}^l v^{in}_i + \sum_{i}^l \val^{in}_i = \sum_{j}^d  v^{out}_j
    \iffull\else\\&\quad\fi \quad\land\quad v^{out}_i \in [0, v_\textsf{max}] \\
    &\}
\end{align*}

\iffull
\item[\alg{IssueBlindCoin($\sk_\chi, \vk, \Gamma, \vec k^{in}, \val_1^{in}, \dots, \val_\ell^{in}$)}{$\widetilde{\vec\sigma}$}]
\else
\item[\ding{118}]\twocolalgnoding{IssueBlindCoin($\sk_\chi, \vk, \Gamma, \vec k^{in}, \val_1^{in}, \dots, \val_\ell^{in}$)}{$\widetilde{\vec\sigma}$}
\fi
The authority $\chi$ parses $\sk_\chi = (x, y_0, y_1, y_2)$, $\vk=(\gamma_0, \gamma_1, \gamma_2, \alpha, \beta_0, \beta_1, \beta_2)$, and $\Gamma = (\vec \sigma'^{in}, \vec \kappa, \cm, \vec \ck,  \vec \cq, \vec \cv, \pi_r)$.
Recompute $\hat{h}_j = H(\cm_{j})$ for each $1 \leq j \leq d$.

\noindent Verify the proof $\pi_{r}$ using $\Gamma$, $\vec h_{*}$, $\vk$, and $\val_1^{in}, \dots, \val_\ell^{in}$. For each $1 \leq i \leq \ell$, parse $\sigma'^{in}_i = (h'_i, s'_i)$, verify $h_i' \neq 1$, and that the following bilinear evaluation holds:
\[
    e(h_i',\kappa_i + \beta_0^{k^{in}_i}) = e(s_i', g_2)
\]
\noindent If one of these checks fail, stop the protocol and output $\perp$.  Otherwise, compute:
\[
\widetilde{s}_j = \hat{h}_j^{x} \ck_{j}^{y_{0}} \cq_{j}^{y_{1}} \cv_{j}^{y_{2}}
\]
and output $\widetilde{\sigma}_j = (\hat{h}_j, \widetilde{s}_j)$.

\item[\alg{PlainVerify($\vk, \sigma, k, q, v$)}{\texttt{true/false}}] \
Parse $\sigma = (h, s)$ and $\vk=(\gamma_0, \gamma_1, \gamma_2, \alpha, \beta_0, \beta_1, \beta_2)$. Reconstruct $\kappa = \alpha \beta_0^k \beta_1^q \beta_2^v$. output \texttt{true} if $h \neq 1$ and $ e(h, \kappa) = e(s, g_2)$; otherwise output \texttt{false}.

\end{description}
The user then calls \textsf{AggCred} and \textsf{Unblind} over each $\widetilde{\sigma}_j$ exactly as described in \Cref{subsec:coconut++}.

\end{document}